\documentclass[pre,preprint,showpacs]{revtex4-1}
\usepackage{bm}
\usepackage{graphicx,color}
\usepackage[export]{adjustbox}
\usepackage{amsmath,amssymb,latexsym}
\DeclareGraphicsExtensions{.jpg,.png}

\usepackage{tikz}
\usetikzlibrary{shapes,arrows}
\tikzstyle{decision} = [diamond, draw, fill=blue!20, 
    text width=3em, text badly centered, node distance=5cm, inner sep=0pt]

\tikzstyle{block} = [rectangle, draw, fill=blue!20, 
    text width=5em, text centered, rounded corners, minimum height=4em]
\tikzstyle{block2} = [rectangle, draw, fill=blue!20, 
    text width=10em, text centered, rounded corners, minimum height=8em]
\tikzstyle{circle} = [circle, draw, fill=blue!20, 
    text width=5em, text centered, rounded corners, minimum height=4em]
\tikzstyle{line} = [draw, -latex']
\tikzstyle{cloud} = [draw, ellipse,fill=red!20, text width=8em, text centered, node distance=3cm,
    minimum height=2em]

\begin{document}

\title{Immersed boundary approach to biofilm spread on surfaces} 
\author{Ana Carpio\cite{carpio:email}}
\affiliation{Departamento de Matem\'atica Aplicada, Universidad Complutense, 
28040 Madrid,  Spain}

\author{Rafael Gonz\'alez}
\affiliation{Departamento de Matem\'atica Aplicada, Universidad Complutense, 
28040 Madrid, Spain \cite{rafael:email}}

\date{\today}


\begin{abstract} We propose a computational framework to study the growth 
and spread of bacterial biofilms on interfaces, as well as the action of antibiotics
on them. Bacterial membranes are represented by boundaries immersed in a fluid matrix and subject to interaction forces. Growth, division and death of bacterial cells follow dynamic energy budget rules, in response to variations in environmental concentrations of nutrients, toxicants and  substances released by the cells. In this way, we create, destroy and enlarge boundaries, either spherical or rod-like. Appropriate forces represent details of the interaction between cells, and the interaction with the environment. Numerical simulations illustrate the evolution of top views and diametral slices of small biofilm seeds, as well as the action of antibiotics. We show that cocktails of antibiotics targeting active and dormant cells
can entirely eradicate a biofilm.
\end{abstract}


\pacs{87.18.Fx, 87.17.Aa, 87.18.Hf, 87.64.Aa}
\maketitle

\section{Introduction} 
\label{sec:intro}

\begin{figure}[h!] \centering
\includegraphics[width=12cm]{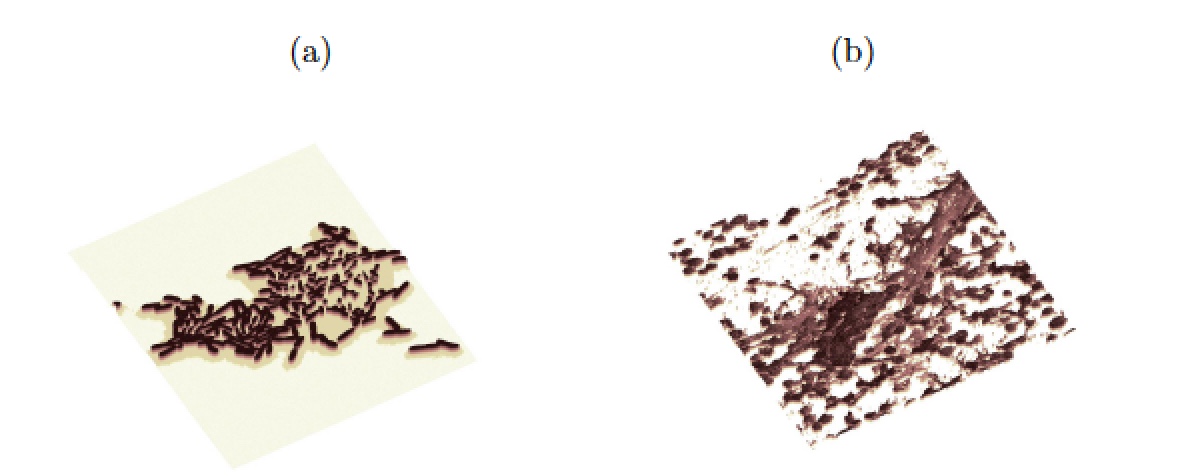}
\caption{Schematic view of the early stages of a biofilm growing on a surface:
Bacteria plus polymeric slime for (a) rod-like bacteria, (b) spherical bacteria.}
\label{fig1}
\end{figure}

\begin{figure}[h!] \centering
\includegraphics[width=7.5cm,valign=t]{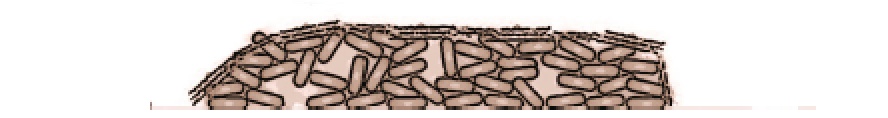}
\caption{Scheme of a vertical slice of a biofilm seed.}
\label{fig2}
\end{figure}

Biofilms are formed by bacteria glued together by a self-produced polymeric
matrix and attached to a moist surface \cite{biofilm}. The polymeric envelop
makes biofilms extremely resistant to antibiotics, disinfectants and chemical 
or mechanical aggressions \cite{Hoiby}. Experiments reveal that their structure 
varies according to environmental conditions. When they grow in flows 
\cite{streamers, Laspidou, Picioreanu_fluids, ibm_deformation}, we see scattered 
bacteria immersed in large chunks of polymer. When they form on interfaces 
with air or tissue, volume fractions of polymer are very small 
\cite{Seminara, Picioreanu_surface, Allen} and biofilms resemble aggregates of 
spherical or rod-like particles, see Figure \ref{fig1} for a view of very early stages.
As they mature, three dimensional sheets are formed, see Figure \ref{fig2}.

Modeling bacterial growth in the biofilm habitat is a complex task 
due to the need to couple cellular, mechanical and chemical processes acting on 
different times scales. Many approaches have been proposed, ranging from purely
continuous models \cite{Seminara} to agent based descriptions \cite{Laspidou, 
Picioreanu_fluids, ibm_deformation,  Picioreanu_surface, Allen} and hybrid 
models combining both \cite{poroelastic,solid/fluid}.
Complexity increases when we aim to take bacterial geometry into account, 
issue that we intend to address here borrowing ideas from immersed
boundary (IB) methods \cite{Peskin02}. 
These methods have already been adapted to simulate
different aspects of biofilms in flows, such as finger deformation  
\cite{ibm_deformation}, attachment of floating bacteria \cite{ibm_adhesion}, 
and viscoelastic behavior   \cite{ibm_rheology}. 
Cell growth and division were addressed by removing the incompressibility 
constraint on the surrounding flow and including `ad hoc' inner sources 
\cite{ibm_division}. Recent
extensions to multicellular growth consider closely packed deformable cells 
attached to each other  \cite{ibm_tumor,ibm_multicellular}. Biofilms growing
on  interfaces differ from multicellular tissues in several respects. First, bacterial 
shapes are more rigid, usually spheres or rods. Second, bacteria remain at a short, 
but variable, distance of each other. To describe their evolution we need to take 
into account at least:
\begin{itemize}
\item Bacterial activities, such as growth, division and death in response
to the environmental conditions.
\item Chemical processes,  such as diffusion of oxygen, nutrients,
and  toxicants (waste products, antibiotics) and production of
autoinducers.
\item Mechanical processes, such as the interaction of the fluid with the 
immersed structures and the interaction between the structures themselves.
\end{itemize}
These processes evolve in different time scales. Compared to cellular
processes, which develop in a time scale of hours, mechanical  and chemical 
processes are quasi-stationary. The inherent time scale for them would be 
seconds. Fast flow processes like adhesion or motion carried by  a flow
are not relevant for biofilms spreading on a surface. Instead, water
absorption from the substrate in the time scale of growth is a factor to 
consider. Variations in the biofilm are driven by cellular activities, in a time 
scale of hours, through changes in the immersed boundaries due to cell 
growth, division, and death \cite{Seminara,Hera,poroelastic}. These
processes are influenced by the secretion of autoinducers and the production 
of waste products and polymers \cite{Seminara,Hera,poroelastic}.

Here, we propose a computational model that combines an IB description of 
cellular arrangements and mechanical interactions with a dynamic energy 
budget representation of bacterial activity and chemical processes, including 
the action of toxicants. Modeling biofilm response to antibiotics is a crucial 
issue in their study \cite{Hoiby}.
The paper is organized as follows. Section \ref{sec:model}
introduces the submodels for the different mechanisms. Section \ref{sec:nond}
nondimensionalizes the equations. Computational issues are discussed
in Section \ref{sec:computational}, while presenting numerical simulations for 
horizontal spread. Section \ref{sec:barrier} considers spread of slices on
barriers. Finally, Section \ref{sec:extinction} shows how biofilm extinction can
be achieved combining two types of antibiotics, one targeting active cells
in the outer layers and another one targeting dormant cells in the biofilm core.
Section \ref{sec:conclusions} summarizes our conclusions.

\section{Model}  
\label{sec:model}

Taking the IB point of view \cite{Peskin95,Peskin02}, we consider the biofilm
as a collection of spherical or rod-like cells, represented by their boundaries,
immersed in a viscous fluid and subject to forces representing interactions,
which are influenced by cell activity as we describe next.  We will formulate
the model in 2D.

\subsection{Immersed boundary representation}
\label{sec:ib}


Let us first describe the basic geometrical arrangement.  To fix ideas, we
consider the schematic structure depicted in Figure \ref{fig1}, a region $\Omega$
containing fluid and bacteria.
We characterize bacteria by immersed boundaries representing their membranes.
We assume the immersed boundaries have zero mass and are permeated by 
fluid. This liquid containing dissolved substances is considered incompressible.
To simplify, we assume that the properties of the liquid are uniform.

The  governing equations are established in \cite{Peskin95,Peskin02}.
We summarize them here, including variations to adapt them to our biofilm
framework:
\begin{itemize}
\item {\it Incompressible Navier-Stokes equations} in $\Omega$ with friction
\begin{eqnarray}
{\partial \mathbf u \over \partial t}
+  \mathbf u \cdot \nabla \mathbf u = \nu \Delta \mathbf u
- {1\over \rho} \nabla p + {1\over \rho} \mathbf f - {\alpha \over \rho} 
\mathbf u,
\quad {\rm div} (\mathbf u) = 0, \label{ib1}
\end{eqnarray}
where $\mathbf u(\mathbf x,t)$ and $p(\mathbf x,t)$ are the fluid 
velocity and pressure, while
$\rho$, $\nu={\mu \over \rho}$ and $\alpha$ stand for the fluid density, 
kinematic viscosity and  friction coefficient, respectively. The source $\mathbf f $ 
represents the force density, that is, force per unit volume.

\item {\it Force spread.} The force $\mathbf f(\mathbf x, t)$ created 
by the immersed boundary (IB) on the fluid is given by
\begin{eqnarray}
\mathbf f(\mathbf x, t)= \int_{\Gamma} \mathbf F(\mathbf q,t)
\delta(\mathbf x- \mathbf X(\mathbf q,t)) \, d \mathbf q,
\label{ib2}
\end{eqnarray}
where $\mathbf X(\mathbf q,t)$ is the parametrization of an immersed
boundary $\Gamma,$ and $\mathbf F(\mathbf q,t)$ the force density 
on it. The integration parameters $\mathbf q$ represent 3D angles.

\item {\it Velocity interpolation.} The evolution equation for the IB
\begin{eqnarray}
{\partial \mathbf X \over \partial t} = \int_{\Omega}
\mathbf u(\mathbf x,t) \delta(\mathbf x- \mathbf X(\mathbf q,t))
\, d \mathbf x  +  \lambda \big( (\mathbf F_g \cdot \mathbf n) 
\mathbf n + \mathbf F_{ext} \big), 
\label{ib3}
\end{eqnarray}
is obtained correcting the no-slip condition with a term representing 
the contribution of the growth forces $\mathbf F_g$ on the IB.
$\mathbf n$ stands for the unit outer vector. Notice that elastic
forces within the IB do not contribute to this term because they are 
tangent to the normal $\mathbf F_e \cdot \mathbf n = 0$. 
$\mathbf F_{ext}$ represents additional external forces that move 
bacteria as blocks, it includes at least interaction forces $\mathbf F_i.$
The adjusting factor $\lambda$ has units ${s \over kg}$.
\end{itemize}
Fluid-structure interaction is mediated by delta functions $\delta$.
In practice, the $\delta$ function is replaced for computational
purposes with approximations which scale with the meshwidth like 
$1/L^3$ in $3D$.  Adequate regularizations are discussed in
\cite{Peskin95,Peskin02}. We locate the immersed boundaries
far from the borders of the computational domain, and enforce
periodic boundary conditions for the fluid on them.

The above equations differ from standard IB models in two respects.
First, we include friction in Navier-Stokes equations (\ref{ib1}) as a way 
to represent  the presence of polymeric threads hindering bacterial
displacement. We could include threads joining the cells as part
of the immersed structures, but we have chosen to represent their
influence through friction in the fluid and interaction forces between
the bacteria, to be described later.
Second, we consider that the forces on the immersed boundaries
are more general than just the elastic forces within it.  This results in
the addition of the term $(\mathbf F_g \cdot \mathbf n) \mathbf n$ in
equation (\ref{ib3}) for their dynamics and allows to connect the
growth forces to a description of cell metabolism.

\subsection{Forces}
\label{sec:forces}

In our case, the IB $\mathbf X$ is composed of many disjoint boundaries
$\mathbf X_j$, $j=1,\ldots, N,$ representing  the membranes of individual
bacteria. The total force density $\mathbf F$ on the IB is the sum of several
contributions.
\begin{itemize}
\item {\it Elastic forces $\mathbf F_e$.} In general, the elastic
forces take the form $\mathbf F_e = - {\partial E \over \partial \mathbf X}$,
where $E(\mathbf X)$ is an elastic energy functional defined on the
immersed boundary configuration $\mathbf X.$

In a two dimensional setting, and assuming the boundary is formed by 
Hookean springs with zero rest length and parametrized by the angle $\theta$,
the force would be
\begin{eqnarray}
\mathbf{F}_{e} = \frac{\partial}{\partial \theta}\left(K\frac{\partial \mathbf{X}}
{\partial \theta}\right), \label{f1}
\end{eqnarray}
for an elastic parameter $K$ (spring constants have units $N/m$). If we 
modify  formula (\ref{ib2}) to calculate a force per unit area $\mathbf f$
\begin{eqnarray}
\mathbf f(\mathbf x, t)= \int_0^{2\pi} \mathbf F(\theta,t)
\delta(\mathbf x- \mathbf X(\theta,t)) \, d \theta, \label{f1b}
\end{eqnarray}
then $\delta$ should include units $1/L^2.$ These forces are calculated
on each component $\mathbf X_j$, $j=1, \ldots, N$.

\item {\it Interaction forces $\mathbf F_i$.}  Bacteria adopt  typically spherical 
({\em coccus}), rod-like ({\em Bacillus}, {\em Pseudomonas}) or spiral ({\em Vibrio}) 
shapes. We focus on the first two types here. Bacteria in a biofilm loose their cilia 
and flagella, that is, their ability to move on their own. On one hand, there are 
repulsive forces between membranes that prevent bacteria from colliding.
On the other, polymeric threads keep bacteria together. As mentioned earlier, we might add a thread network. However,  we choose to represent their action by means of a friction term in Navier-Stokes equations. In this way,  we avoid adding thread networks to keep cells together. We just need to separate the cells as they grow or divide.

When the distances between bacteria are below a critical distance, 
repulsion forces act fast. The repulsion force $\mathbf F_{i,j}$ acting on each 
bacterium with boundary $\mathbf X_j$, $j=1,...,N$, depends on the distance 
between all pairs. For spherical  bacteria, we set the force as follows:
\begin{eqnarray} \begin{array}{l}
\mathbf F_i = \sum_{j=1}^N \mathbf F_{i,j} \delta_j, \\
\mathbf{F}_{i,j} = \left \{ \begin{matrix} \displaystyle \sum_{n=1, n\neq j}^{N} 
\frac{\sigma}{d_{min}} \mathbf{n}_{{\rm cm},n,j} & \mbox{if }d_{j,n} \leq d_{min},
\\ 
\displaystyle\sum_{n=1, n\neq j}^{N} {\sigma\left(1+\tanh\left({s_p-d_{j,n} \over v_p}\right)\right) 
\over 2d_{j,n}} \mathbf{n}_{{\rm cm},n,j} & \mbox{if }d_{j,n} > d_{min}, \end{matrix}\right. 
\end{array}
\label{f2} 
\end{eqnarray}
where $\sigma$ is the repulsion parameter with appropriate units, $d_{j,n}$ is the smallest distance between the curves defining bacteria $j$ and $n$, $N$ is the number of bacteria, and
$\mathbf n_{{\rm cm},n,j} = {\mathbf X_{c,j} - \mathbf X_{c,n} \over \| \mathbf X_{c,j} - \mathbf X_{c,n} \|}$ is the unit vector that joins the centers of mass, oriented from $n$ to $j$.
Here, $\delta_j$ takes the value $1$ at the nodes of the cell boundary $\mathbf X_j$ and vanishes on other cell boundaries. Additional parameters govern
the minimum value $d_{min}$ that $d_{j,n}$ can take, the order of magnitude of 
this force $s_p$, and the decay as the distance decreases $v_p$.
These forces are similar for spheres and rods, changing the parameter values, 
see  Table \ref{table1}.

\item {\it Growth forces $\mathbf F_g$.} 
Growth of spherical bacteria is described through 
variations in their radius, whereas rod-like bacteria grow in length.  
Assuming the rate of growth of their radius (resp. lengths) are
known, the effect on each cell boundary would be, for spheres, 
\begin{eqnarray}
 \frac{dR_j}{dt} {\mathbf X_j- \mathbf X_{c,j} \over 
 \|\mathbf X_j - \mathbf X_{c,j} \|}  = \frac{dR_j}{dt} \mathbf n, 
\quad j=1, \ldots, N, \label{f3sphere}
\end{eqnarray}
where $R_j$ and $\mathbf X_{c,j}$ denote the radius and center  
of the bacterium $\mathbf X_j$. For rods, growth forces act on the edges,
forcing a change of length
\begin{eqnarray}
 {1\over 2} \frac{dL_j}{dt} \mathbf l, \quad j=1, \ldots, N, \label{f3rod}
\end{eqnarray}
where $\mathbf l$ is an outer unit vector along the rod axis. Notice
that for spheres 
$(\frac{dR_j}{dt} \mathbf n \cdot \mathbf n) \mathbf n = \frac{dR_j}{dt} \mathbf n$
whereas for rods
$(\frac{dL_j}{dt} \mathbf l \cdot \mathbf n) \mathbf n \sim 0$
except on the rod edges. We take $\mathbf F_g$ proportional
to these growth factors.

Our description of cell metabolism in Section \ref{sec:deb} provides the 
required equations for the time dynamics of radii $R_j$ and lengths $L_j$.
\end{itemize}

Finally, the total force we have to spread to the fluid through (\ref{ib2})
or (\ref{f1b}) is the sum of all the 
forces  $\mathbf F = \mathbf F_e - \mathbf F_i + \mathbf F_g$.

\subsection{Cellular activity}
\label{sec:deb}

We  describe bacterial metabolism by means of a dynamic
energy budget approach \cite{Deb18,DebBook,Debparameter}:

\begin{itemize}
\item {\it Dynamic energy budget equations for cell metabolism.}
Bacteria transform nutrients and oxygen in energy, which they use for 
maintenance, growth and division. In a biofilm, some cells undergo phenotypical 
changes and start performing new tasks. For instance, some become
producers of exopolysaccharides, that is, the extracellular polymeric
substances forming the biofilm EPS matrix. This is more likely for cells 
with scarce resources \cite{Hoiby,Hera} to sustain normal reproduction 
and growth.

Given an aggregate formed by $N$ bacteria, their energy $e_j$ and
volume $V_j$, $j=1,...,N,$ evolve according to
\begin{eqnarray} 
\frac{de_j}{dt} &= \nu' \left(\frac{S}{S + K_S} - e_j\right), &
\nu' = \nu e^{- \gamma \varepsilon}  \left( 1 + {C_{out} \over K_v} 
\right)^{-1},  \label{deb1} \\
\frac{dV_j}{dt}& = \left(r_j \frac{a_j}{a_M} - h_j\right) V_j, &
\quad r_j = \left(\frac{\nu'e_j -m g}{e_j + g}\right)^+, \label{deb2}
\end{eqnarray}
where $\nu$ is the energy conductance, $\nu'$ the conductance
modified by exposure to a toxicant, $m$ the maintenance rate, $g$ the 
investment ratio, $a_M$ the target acclimation energy, $K_S$ a half-saturation
coefficient, $K_V$  the noncompetitive  inhibition coefficient and 
$\gamma$ the environmental degradation effect coefficient. 
The factor $r_j$  denotes the bacterial production rate. The symbol 
$^+$ stands for `positive part', which becomes zero for negative values.
The variables $S$, $C_{out}$, $\varepsilon$ denote the limiting 
nutrient/oxygen concentration, the concentration of toxic products, 
and the environmental degradation, respectively. Note that, for
spherical bacteria with radius $R_j$, we have $V_j= {4 \over 3} \pi R_j^3$.
In 2D, $V_j= \pi R_j^2$,  and (\ref{deb2}) implies
\begin{eqnarray}
 2 \frac{dR_j}{dt}  = \Big(r_j \frac{a_j}{a_M} - h_j\Big) R_j. \quad
\label{debradius}
\end{eqnarray}
For rod-like bacteria of radius $R$ and length $L_j$, $V_j\sim \pi R^2 L_j$. 
In 2D, $V_j\sim 2R L_j.$ For ellipsoidal approximations,  
$V_j = \pi b L_j,$ where $b$  is the small and $L_j$ the great semi-axes, with
\begin{eqnarray}
\frac{dL_j}{dt} = \Big(r_j \frac{a_j}{a_M} - h_j\Big) L_j. \label{deblength}
\end{eqnarray}

These equations must be complemented with equations for
cell response  to the  degradation of the environment and the 
accumulation of toxicants. The cell  undergoes damage,
represented by aging $q_j$ and hazard $h_j$ variables, as well
as acclimation, represented by the variable $a_j$. 
For $j=1,\ldots, N$, these additional variables are governed by
\begin{eqnarray}
\frac{dq_j}{dt} = e_j (s_G \rho_x {V_j \over V_T} q_j + h_a) (\nu' - r_j) + 
k_{tox} C_{in,j} - (r_j + r_{e,j}) q_j, \label{deb3} \\
\frac{dh_j}{dt} = q_j - (r_j + r_{e,j}) h_j, \label{deb4} \\
\frac{dp_j}{dt} = - h_j p_j, \label{deb5} \\
\frac{dC_{in,j}}{dt} = k_{in} C_{out} - k_{out} C_{in,j} - (r_j + r_{e,j}) C_{in,j}, 
\label{deb6} \\
\frac{da_j}{dt} = (r_j + r_{e,j}) \left(1 - \frac{a_j}{a_M} \right)^+, \label{deb7}
\end{eqnarray}
where $\rho_x$ is the cell density, $s_G$ a multiplicative stress coefficient, $h_a$ 
the Weibull aging acceleration,  and $k_{tox}$, $k_{in}$, $k_{out}$ the toxicity, influx 
coefficient and  efflux coefficient of toxicants, respectively. The variable
$C_{in,j}$ denotes the toxicant cellular density inside the cell
and $p_j$ its probability of survival at time $t$. The factor $r_{e,j}$ is non zero
only when the cell is an EPS producer (the values of the parameters $m$ and $g$ 
may be slightly different for such cells). In that case the rate of EPS production 
$r_{e,j} = k r_j + k'$, where $k$ is the growth associated yield whereas $k'$ is the 
non growth associated yield. The produced EPS is then
\begin{eqnarray}
{d V_{e,j} \over dt} = r_{e,j} V_{j}. \label{deb8}
\end{eqnarray}
A fraction $\eta$ of the produced EPS stays around the cell, while
a fraction $\eta \in (0,1)$ diffuses taking the form of a concentration
of monomers $C_e$.

\item {\it Equations for concentrations.}
System (\ref{deb1})-(\ref{deb7}) describes the metabolic state of each bacterium,
and is coupled to reaction-diffusion equations for the relevant concentrations
in $\Omega$:
\begin{eqnarray}
\frac{dS}{dt} = - \nu' \frac{S}{S + K_S} \rho_x \displaystyle\sum_{j} {V_j\over V_T} 
\delta_j + d_s \Delta S - \mathbf{u} \cdot \nabla S, \label{c1} \\
\frac{dC_e}{dt} = \eta \rho_x \displaystyle\sum_{j} r_{e,j} {V_j\over V_T} \delta_j
+ d_e \Delta C_e - \mathbf{u} \cdot \nabla C_e, \label{c2} \\
\frac{dC_{out}}{dt} = - C_{out} \displaystyle\sum_{j} r_j \delta_j 
+ d_c \Delta C_{out} - \mathbf{u} \cdot \nabla C_{out}, \label{c3} \\
\frac{d\varepsilon}{dt} = \nu_{\varepsilon} \rho_x \displaystyle\sum_{j} 
(r_j + \nu_m m){V_j\over V_T} \delta_j + d_{\varepsilon} \Delta \varepsilon 
- \mathbf{u} \cdot \nabla \varepsilon, 
\label{c4}
\end{eqnarray}
where $\nu_{\varepsilon}$ is the environmental degradation coefficient, $\nu_m$ is 
the maintenance respiratory coefficient and $d_{\varepsilon}$, $d_s$, $d_e$, $d_c$  
the diffusion coefficients for degradation $\varepsilon$, limiting oxygen/nutrient concentration $S$, monomeric EPS $C_e$, and toxicants $C_{out}$, respectively. Here $\delta_j$ equals one in the region  occupied by cell $j$, vanishes otherwise. $V_T$ is a reference volume.
These equations are typically solved in the computational domain with no flux boundary  conditions, except for $S$, which has a constant supply at the borders, and $C_{out}$ which is supplied at the borders as prescribed.

\item {\it Spread of cellular fields and interpolation of concentration fields.}
The system of ordinary differential equations (\ref{deb1})-(\ref{deb8}) and 
reaction-diffusion equations (\ref{c1})-(\ref{c4}) are coupled using a similar
philosophy as that in IB models. However, now we transfer information
not between curves and a two dimensional region but between confined
regions occupied by bacteria and the whole computational domain:
\begin{itemize}
\item Spread of fields defined on bacteria:
Equations (\ref{c1})-(\ref{c4}) use the cell volumes and rates as sources
and sinks for the concentrations.
\item Interpolation of global fields on the bacteria: 
For each bacterium, system (\ref{deb1})-(\ref{deb8}) uses the averaged
values of $S$, $C_{out}$, $\varepsilon$ in the region occupied by the cell.
$C_{out}$ represents the dissolved (extracellular) concentration of toxicants.
\end{itemize}

\end{itemize}

\section{Nondimensionalization of the equations} 
\label{sec:nond}

For computational purposes, it is essential to nondimensionalize properly these sets of equations. This allows us to identify relevant time scales for the different sets of equations, as well as controlling parameters.
To remove dimensions we have to choose characteristic values for the different
magnitudes. The characteristic length $L$ will tell us what part of the problem we want to focus  on, that is, if we prefer to study what happens with the whole set of bacteria and do not want to spend a lot of computational time solving for details, or if we want to give more importance to what happens in the smaller areas. In our case we are interested in small cell aggregates, so we will have a characteristic length of $L=10 [\rm \mu m]$ (microns, $1 \mu$m $= 10^{-6}$m), because it is about the maximum length of rod-like bacteria. In general, it will be the size of a small group of them. Time scales vary: microseconds for fluid processes, seconds for diffusion processes, and hours for cellular processes. 

Let us first consider the IB submodel. We set a characteristic time $T = 10^{-6}$[s]. In equation (\ref{ib1}), the terms $(u_t + u \nabla u), \nu \Delta u$ have the same units, regardless of dimension. Let us set $p'= {p\over \rho}$, $\alpha'={\alpha \over \rho}$. Then, $\mathbf f' ={\mathbf f\over \rho}$ has units of acceleration. Formally, one can just suppress one dimension in the variables and derivatives and use in 2D:
\begin{eqnarray}   {\partial \mathbf u \over \partial t}
+  \mathbf u \cdot \nabla \mathbf u = \nu \Delta \mathbf u
- \nabla p' + \mathbf f '- \alpha' \mathbf u.  \label{aux1}
\end{eqnarray}
As a reference acceleration, we set $a_0 = {E \over \rho L} = {E_s \over \rho_s L}$, where $E_s$ is a longitudinal tension in units [${{\rm N} \over {\rm m}}$] (Young modulus for springs) and $\rho_s$ surface density in units [${{\rm kg} \over {\rm m}^2}]$.
We know 3D values for the parameters. The Young modulus $E$ for bacterial membranes \cite{Youngcell1} lies in the range $100-200$ [MPa]. We set $E=150$ MPa = $150 \times 10^6 $ [${{\rm N} \over {\rm m^2}}$].  The density of water/biomass $\rho$  \cite{Seminara} is about $10^3$  $[{{\rm kg} \over {\rm m}^3}]$. In this way, we find a value for $a_0$.
Regarding the forces (\ref{ib2}), for the elastic contribution we use (\ref{f1}) and (\ref{f1b})  in 2D, which relates force per unit area to force with $\delta$ in units of ${1 \over L^2}.$ 

\begin{table}[ht!]
\begin{center} \begin{tabular}{|c|c|c|c|}
 \hline
Name & Symbol & Values & Units     \\ \hline
Biomass density & $\rho$ & $10^3 $ & $[\rm kg/m^3]$   \\ \hline
Biomass viscosity & $\mu$ & $100 $ & $[\rm kg/(m\, s)]$  \\ \hline
Bacterial membrane Young Modulus & $E$ & $150 \times 10^6 $ 
& $[\rm kg/(m\, s^2)]$  \\ \hline
\end{tabular}
\end{center} \vskip -3mm
\caption{Values for dimensional parameters of the IB submodel expressed in their standard units.
}
\label{table1}
\vskip 5mm
\end{table}

\begin{table}[ht!]  
\begin{center} \begin{tabular}{|c|c|c|c|c|c|}
 \hline
\!\! $t\!=\!T \tilde t_1$ \!\!&\!\! $x \!=\! L \tilde x$ \!\!&\!\!  
$u \!=\! U \tilde u$ \!\!&\!\!  $ {p \over \rho} \!=\! 
P \tilde p$ \!\!&\!\! $\mathbf F \!=\! F  \tilde{ \mathbf F} $ \!\!&\!\! 
$ {\mathbf f \over \rho} \!=\!{\mathbf f_s \over \rho_s} \!=\! a_0 \tilde{ \mathbf f} $ 
\!\!  \\ \hline
\!\!  $\delta \!=\! {1\over L^2} \tilde \delta$ \!\!&\!\! $K \!=\! K_0 E_s$ 
\!\!&\!\! $U\!=\!{L \over T}$  \!\!&\!\!  $P\!=\!{L^2 \over T^2}$ \!\!&\!\! $F\!=\! E_s L$  
\!\!&\!\!  $a_0 \!=\!{E \over L \rho} \!=\! {E_s \over L \rho_s } $ 
\!\!  \\ \hline
\!\!  $\alpha\!=\! \alpha_0 {\rho \over  T}$ \!\!&\!\!  $\lambda\!=\! {\lambda_0 \over  E_sT}$ 
\!\!&\!\! $d_{j,n} \!=\! L \tilde d_{j,n}$ \!\!&\!\!  $\sigma \!=\! \sigma_0 E_s L^2 $ 
\!\!&\!\!  $s_{p} \!=\! s_{p,0} L$  \!\!&\!\! $v_{p} \!=\! v_{p,0} L$   
\!\! \\ \hline
\end{tabular} \end{center} \vskip -3mm
\caption{Change of variables used to nondimensionalize the IB equations. The
$\tilde{} $ symbols are dropped  for ease of notation  after it. Dimensionless
parameters  $K_0$, $\alpha_0$, $\sigma_0$, $\lambda_0$, $s_{p,0}$,
$v_{p,0}$, as well as the dimensionless numbers $Re$, $F_c$ and
dimensional values for $\rho$, $\mu$, $E$ are given in
Tables \ref{table1} and \ref{table3}. The unknown value $E_s$ scales out.
We assume $E/\rho = E_s/\rho_s$.}
\label{table2}
\vskip 5mm
\end{table}

\begin{table}[ht!]
\begin{center} \begin{tabular}{|c|c|c|c|c|c|c|c|c|c|c|c|c|}
 \hline
$Re={\rho L^2 \over \mu T}$ & $F_c = {T^2 E \over L^2 \rho}$  &
$\alpha_0$ & $\lambda_0$ & $K_0$ & $\sigma_0$ & 
$d_{min,0}$ & $s_{p,0} $  & $v_{p,0} $ 
 \\ \hline
$10^{-3}$ & $1.5 \times 10^3$ & 
$10^{-3}$ & $10^4$ & $0.15$ & $0.05$ &
0.01 & $0.01$ & $0.01$
 \\ \hline
\end{tabular} \end{center} \vskip -3mm
\caption{Dimensionless control parameters for the IB submodel (\ref{ib1ad})-(\ref{ib6ad}) when $L= 10^{-5}$ [m] and $T=10^{-6}$ [s].
}
\label{table3}
\end{table}

Performing the changes of variables indicated in Table \ref{table2} and dropping
the $\tilde{}$ symbol for ease of notation we find the dimensionless IB system
with parameters given by Tables \ref{table1}-\ref{table3}:
\begin{eqnarray}  
 {\partial \mathbf u \over \partial t_1} \!+\!  \mathbf u \!\cdot\! \nabla \mathbf u 
\!=\! {1\over Re} \Delta \mathbf u \!-\!  \nabla p 
\!+\!  F_c \mathbf f  \!-\! \alpha_0 \mathbf u,
\; {\rm div}(\mathbf u) \!=\! 0, \label{ib1ad} \\
\mathbf f(\mathbf x, t_1)= \int_0^{2\pi} \mathbf F(\theta,t_1)
\delta(\mathbf x- \mathbf X(\theta,t_1)) \, d \theta, \;  \mathbf X = \cup_{j=1}^N
 \mathbf X_j, \label{ib2ad} \\
{\partial \mathbf X \over \partial t_1} = \int_{\Omega}
\mathbf u(\mathbf x,t_1) \delta(\mathbf x- \mathbf X(\mathbf q,t_1))
\, d \mathbf x  + \lambda_0 \big( (\mathbf F_g \cdot \mathbf n) 
\mathbf n  + \mathbf F_{ext} \big),
\label{ib3ad} \\
\mathbf F = \mathbf F_e + \mathbf F_g - \mathbf F_i,  \label{ib4ad} \\
\mathbf F_e = \frac{\partial}{\partial \theta}\left(K_0 \frac{\partial \mathbf{X}}
{\partial \theta}\right), \; \mathbf F_{ext} = \mathbf F_{i}, \label{ib5ad} \\
\mathbf F_i = \left \{ \begin{matrix} \displaystyle\sum_{j=1}^N\displaystyle \sum_{n=1,n \neq j}^{N} \frac{\sigma_0 \delta_j}{d_{min,0}} \mathbf{n}_{{\rm cm},n,j} & \mbox{if }d_{j,n} \leq d_{min,0},
\\ \displaystyle\sum_{j=1}^N\displaystyle\sum_{n=1, n\neq j}^{N} {\sigma_0\left(1+\tanh\left({s_{p,0}-d_{j,n} \over v_{p,0}}\right)\right)\delta_j \over 2d_{j,n}} \mathbf{n}_{{\rm cm},n,j} & \mbox{if }d_{j,n} > d_{min,0}. \end{matrix}\right. \label{ib6ad}
\end{eqnarray}

The growth term $\mathbf F_g$ would be noticeable in the time scale of hours.
In this scale, it is negligeable.  The effect of growth would come through the boundaries, which move in the time scale of hours due to cellular processes. 
Here $\lambda_0 \mathbf F_g \sim {T \over 3600  } \sim 10^{-10}$. We can
remove $\mathbf F_g$ from these equations. The effect of cell metabolism
on bacterial boundaries will be calculated directly from the DEB equations.

Next, we consider the DEB equations for each cell.
Recall that the variables $e, a, V_j, p_j, C_{in}$ are dimensionless.
Hazard $h$ and aging $q$ have units $\rm hour^{-1}$ and $\rm hour^{-2}$, respectively.
We remove the dimensions in the variables as indicated in Table \ref{table4}. 
Taking into account the parameter values listed in Table \ref{table5},
the remaining dimensions for parameters and rates are consistent. We
work in a timescale $\tau = 1$ hour, which is the natural step. Dropping again  
the symbol $\, \tilde{} \, $  for ease of notation we find for each cell $j$
\begin{eqnarray} 
\frac{de_j}{d t_2} &= \tau \nu' \left(\frac{S}{S + 1} - e_j\right), &
\nu' = \nu e^{- \gamma \varepsilon}  \left( 1 + {C_{out} K_S\over K_v} 
\right)^{-1},  \label{deb1ad} \\
\frac{dV_j}{d t_2}& = \left(\tau r_j \frac{a_j}{a_M} - h_j\right) V_j, &
\quad r_j = \left(\frac{\nu'e_j -m g}{e_j + g}\right)^+, \label{deb2ad} 
\end{eqnarray}
and
\begin{eqnarray}
\frac{dq_j}{d t_2} = e_j (s_G \rho_x V_j q_j + h_a \tau^2 ) \tau (\nu' - r_j) + 
\tau^3  k_{tox} C_{in,j}  - \tau (r_j + r_{e,j}) q_j, \label{deb3ad} \\
\frac{dh_j}{d t_2} = q_j - \tau (r_j + r_{e,j}) h_j, \label{deb4ad} \\
\frac{dp_j}{d t_2} = - h_j p_j, \label{deb5ad} \\
\frac{dC_{in,j}}{d t_2} = \tau k_{in} K_s C_{out} - 
\tau k_{out} C_{in,j} - \tau (r_j + r_{e,j}) C_{in,j}, \label{deb6ad} \\
\frac{da_j}{d t_2} = \tau (r_j + r_{e,j}) \left(1 - \frac{a_j}{a_M} \right)^+.
\label{deb7ad} 
\end{eqnarray}
For round bacteria in 2D, $V_j= \pi R_j^2$. Equation (\ref{deb2ad})
provides the evolution of $\frac{dR_{j}}{d t_2}$. The evolution
of the boundary due to cell metabolism is given by
\begin{eqnarray}
\frac{\partial \mathbf{X_j}}{\partial t_2} = \frac{dR_{j}}{d t_2} {\mathbf 
X_j(\mathbf q,t_2) - \mathbf X_{c,j}(t_2)
\over \|\mathbf X_j(\mathbf q,t_2)- \mathbf X_{c,j}(t_2) \|}. \label{deb8ad}
\end{eqnarray}
In a similar way, if the cell is rod-like, its boundary evolves as given by
(\ref{deblength}).

\begin{table}[ht!]
\begin{center} \begin{tabular}{|c|c|c|c|c|c|}
 \hline
$t = \tau t_2$ & $h_j  = \tilde h_j  \tau^{-1}$  & $q_j  = \tilde q_j  \tau^{-2}$
& $V_j  = \tilde V_j L^{2}$ & $C_{out}  = \tilde C_{out} K_S$  
& $V_T = L^2$
 \\ \hline
\end{tabular} \end{center} \vskip -3mm
\caption{Change of variables for nondimensionalization of the DEB
model. We set $\tau = 1$ [hour].
}
\label{table4}
\end{table}

\begin{table}[ht!]
\begin{center} \begin{tabular}{|c|c|c|}
 \hline
Symbol & Values & Units       
 \\ \hline
$\nu$ & $0.84768$ & $[\rm hour^{-1}]$ 
\\ \hline
$\gamma$ & $1$ & $[\rm n.d.]$ 
\\ \hline
$K_V$ & $154.82$ & $[\rm mg/l]$ 
\\ \hline
$K_S$ & $0.1$ & $[\rm mg/l]$ 
\\ \hline
$C_{out}$ & $0.20, 0.78, 1.56, 3.13$ & $[\rm mg/l]$ 
\\ \hline
$g$ &  0.9766 &  $[\rm n.d]$ 
\\ \hline
$m$ & $0.1266$ & $[\rm hour^{-1}]$ 
\\ \hline
$\nu_m$ & $0.054703$ & $[\rm n.d.]$ 
\\ \hline
$a_M$ & $1.6703$ & $[\rm n.d.]$  
\\ \hline
$s_G$ & $0.8921 \cdot 10^{-5}$ & $[\rm l/mg]$ 
\\ \hline
$h_a$ & $1.4192 \cdot 10^{-4}$ & $[\rm hour^{-2}]$  
\\ \hline
$\nu_{\varepsilon}$ & $0.23566/12000$ & $[\rm l/mg]$ 
\\ \hline
$k$ & $2.2371$ & $[\rm mg_{polymer}/ mg_{cell}]$ 
\\ \hline
$k'$ & $0.29$ & $[\rm mg_{polymer}/ (mg_{cell} hour)]$  
\\ \hline
$\eta$ & $1/2$ & $[\rm n.d.]$
\\ \hline
$k_{out}$ & $0.17251$ & $[\rm hour^{-1}]$ 
\\ \hline
$k_{in}$ & $5.16 \times 10^{-4}$ & $[\frac{\rm l}{\rm mg \, hour}]$ 
\\ \hline
$k_{tox}$ & $5.416 \times 10^{3}$ & $[\rm hour^{-3}]$ 
\\ \hline
$\rho_x$ & $47000$ & $[\rm mg/l]$  
\\ \hline
\end{tabular}
\end{center}
\caption{Parameters of the DEB submodel (\ref{deb1ad})-(\ref{deb7ad})
expressed in their standard  units, adapted from \cite{Debparameter} and \cite{Deb18}. Note that $ [mg/l] = [10^{-6} kg/10^{-3} m^3]$. 
When inserted in the equations, all must be written in the same units of choice. Special attention must be paid to time units, which will be either hours or seconds, which requires multiplying or dividing by $3600.$}
\label{table5}
\end{table}

Finally, let us consider next the diffusion problems.
The variable $\varepsilon$ is dimensionless.  The concentrations $S$,
$C_e$, $C_{out}$ have units $mg/l$. We set $C= K_S \tilde C$
for all the concentrations, $t = T_d \tilde t_d$ and same spatial scaling as before, 
as indicated in Table \ref{table6}. Removing  $\, \tilde{}\, $  
for ease of notation again, we find the dimensionless equations:
\begin{eqnarray}
\frac{dS}{dt_d} = - T_d \nu'^* \frac{S}{S + 1} {\rho_x  \over K_S} \displaystyle\sum_{j} V_j \delta_j 
+ d_{s,0}D_c \Delta S - {T_d \over T} \mathbf{u} \!\cdot\!  \nabla S, \label{c1ad} \\
\frac{dC_e}{dt_d} = \eta {\rho_x  \over K_S} \displaystyle\sum_{j} T_d r_{e,j}^* V_j \delta_j + 
d_{e,0}D_c    \Delta C_e - {T_d \over T} \mathbf{u} \!\cdot\!  \nabla C_e, \label{c2ad} \\
\frac{dC_{out}}{dt_d} = - C_{out} \displaystyle\sum_{j} T_d r_j^* \delta_j 
+ d_{c,0}D_c  \Delta C_{out} - {T_d \over T} \mathbf{u} \!\cdot\! \nabla C_{out}, \label{c3ad} \\
\frac{d\varepsilon}{dt_d} = \nu_{\varepsilon} \rho_x  T_d \! \displaystyle\sum_{j} 
(r_j^* \!+\! \nu_m m^*)V_j  \delta_j  + d_{\varepsilon,0}D_c  \Delta \varepsilon 
- {T_d \over T} \mathbf{u} \!\cdot\! \nabla \varepsilon, \label{c4ad} 
\end{eqnarray}
with parameters given in Tables \ref{table5} and \ref{table7}.
Notice that $\nu'$, $m$, $r_j$ and $r_{e,j}$ have units $\rm hour^{-1}$. To be used in these equations, they have to be expressed in units of $s^{-1}$, that is, divided by $3600$. We denote  those values by $\nu'^*$, $m^*$, $r_j^*$ and $r_{e,j}^*$.
The new diffusion coefficients will be  large, the same as ${1\over Re}$.
Both  systems for fluids and concentrations should relax fast to an equilibrium.  
We are interested in stationary solutions, to be more precise, quasi-stationary, in the sense that they change with time when the immersed boundaries grow/split/die or the sources vary. That happens in a much longer time scale of hours.

\begin{table}[ht!]  
\begin{center} \begin{tabular}{|c|c|c|c|c|c|}
 \hline
\!\! $t\!=\! T_d \tilde t_2$ \!\!&\!\! 
$x \!=\! L \tilde x$ \!\!&\!\!
$V_j \!=\! L^2 \tilde V_j$ \!\!&\!\!
$S \!=\! K_S \tilde S$ \!\!&\!\! 
$C_e \!=\! K_S \tilde C_e$ \!\!&\!\! 
$C_{\rm out} \!=\! K_S \tilde C_{\rm out}$ 
\!\!  \\ \hline
$V_T = L^2 $ \!\!&\!\! 
$D_c \!=\! D {T_d \over L^2}$ \!\!&\!\! 
$\nu'^* = {\nu \over 3600}$ \!\!&\!\! 
$m^* = {m \over 3600}$ \!\!&\!\! 
$r_j^* = {r_j \over 3600}$ \!\!&\!\! 
$r_{e,j}^* = {r_{e,j} \over 3600}$ 
\!\!  \\ \hline
\end{tabular} \end{center} \vskip -3mm
\caption{Change of variables used to nondimensionalize the equations
for concentrations. The $\tilde{} $ symbols are dropped  for ease of notation  
after it.  }
\label{table6}
\end{table}

\begin{table}[ht!]  
\vskip -3mm
\begin{center} \begin{tabular}{|c|c|c|c|c|}
 \hline
\!\!  
$d_{\varepsilon,0}  \!=\! 22$ \!\!&\!\!  
$d_{s,0} \!=\! 10 $ \!\!&\!\! 
$d_{c,0} \!=\! 5 $ \!\!&\!\! 
$d_{e,0} \!=\! 5 $ \!\!&\!\! 
$D_c =  10$  
\!\!  \\ \hline
\end{tabular} \end{center} \vskip -3mm
\caption{Dimensionless parameters used in the concentration
submodel (\ref{c1ad})-(\ref{c4ad}) when $T_d = 1$ [s] and $D = 10^{-9} $ 
[m$^2$/s].}
\label{table7}
\end{table}

\section{Computational model for unconstrained spread} 
\label{sec:computational}

As said earlier, we are interested in two kinds of two dimensional reductions 
of three dimensional geometries representing biofilms spread on a surface.
The first one consists of top views of early biofilm stages, see Figure \ref{fig1}.
In the second one, we consider a 2D diametral slice of a 3D biofilm, see Figure
\ref{fig2}. Let us focus on the first one, which can be handled with the equations
and nondimensionalizations summarized in the previous two sections. The 
second one requires additional details that we will explain later.

\begin{figure} \centering
\includegraphics[width=12cm]{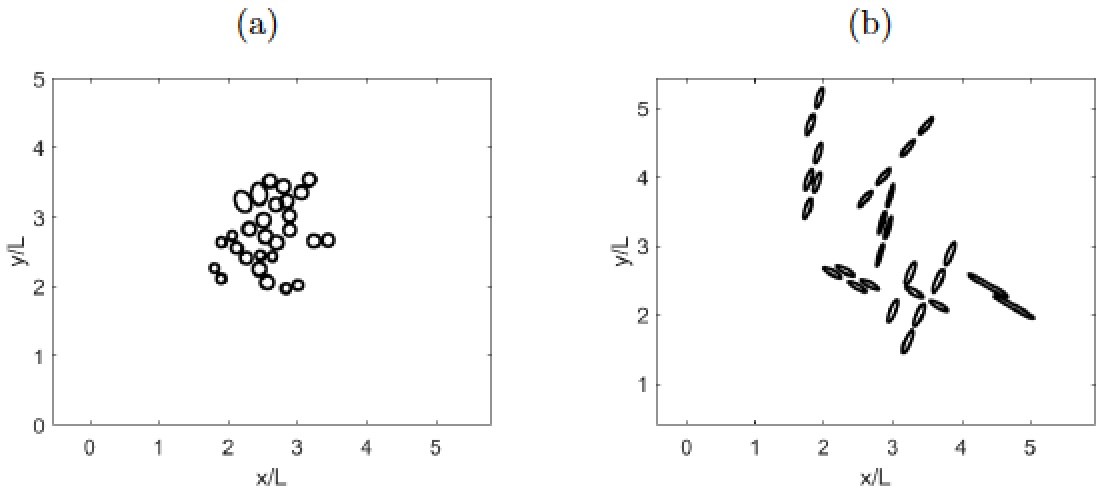}
\caption{Snapshots of the evolution of a few bacteria with initial random sizes varying in the  ranges specified in the text. Clusters formed by (a) round and (b) rod-like bacteria, see Video1 and Video2.
}
\label{fig3}
\end{figure}

To fix ideas, we consider that the computational region has a reference size 
around  $100 \times 100$ [$\rm \mu m$], that is, $10 L \times 10 L$ when 
$L= 10$ [$\rm \mu m$]. We place a few bacteria at the center of this region, and let nutrients and toxicants diffuse from the boundaries.
While a biofilm spreads on an interface with air, bacteria barely move, except
when pushed by the rest. They grow up to their division or shrink until their death. 
The bacterial cluster tends to spread in the direction of the nutrient/oxygen 
concentration gradient. As they divide, bacteria occupy the free space and 
remain at a small distance from their neighbors. 
The average diameter of spherical bacteria is about $0.5-2.0$ [$\rm \mu$m]. 
For rod-shaped or filamentous bacteria, the average length is about $1-10$ 
[$\rm \mu$m] and diameter is about $0.25-1.0$ [$\rm \mu m$].  In our simulations 
we have taken for spheres $R = 0.025 - 0.1$ [$\rm \mu m$], and for rods diameter 
$0.05 - 0.1$ [$\rm \mu m$] and length $ 0.1 - 1$ [$\rm \mu m$], nondimensionalized divided by the reference length $L$.
Figure \ref{fig3}  illustrates some simulations.

Once we have fixed an initial bacterial arrangement and set initial conditions for all the variables we distinguish three blocks of equations.
The DEB equations for each cell (\ref{deb1ad})-(\ref{deb8ad}) are solved in
the time scale of hours.  In that time scale, the IB equations (\ref{ib1ad})-(\ref{ib6ad})
and the equations for chemical processes (\ref{c1ad})-(\ref{c4ad}) are quasistationary, changes are induced by growth, division, or destruction or boundaries according to the DEB submodel and the criteria for division, death or interaction. We solve them using time relaxation, that is, we solve the time dependent problems until the solutions relax to a stationary state. 
More precisely, we proceed as follows.
First, we integrate the DEB system for all cells. Then, we relax the Ib2d model with 
interaction force to a stationary state, and finally the concentrations relax to their
stationary state in the diffusion time scale.
The process is schematized in Flowcharts \ref{fig4} and \ref{fig8}.
We next give details about the discretization and the initialization procedures.

\subsection{Discretization}
\label{sec:discretization}

We define in the computational region  a square mesh
$\mathbf x_{i,j} = (x_i, y_j)$, $i,j=0,...,{\cal N}$, with
step $dx = dy = h$ and nodes $x_i = x_0+ i dx$, $y_j= y_0 + j dy$, where
$x_0 = y_0 = 0$, $x_{\cal N}= y_{\cal N} = {\cal L}.$ We keep this mesh
for all the submodels. However, the three submodels use different time
discretizations.  The main time mesh is $t\ell = \ell dt$, $\ell = 0,..., {\cal M}$, 
up to the final time ${\cal T}= {\cal M} dt.$ For each cell, the systems of ordinary 
differential equations (\ref{deb1ad})-(\ref{deb8ad}) are discretized by a classical 
Runge-Kutta scheme on that mesh with step $dt = 0.01.$
For the other two submodels we seek stationary solutions. We use the time 
dependence to implement time relaxation schemes to approximate them with
adapted time steps.
The reaction-diffusion equations (\ref{c1ad})-(\ref{c4ad}) are discretized by
classical explicit  finite difference schemes.
The whole set of equations for the immersed boundaries (\ref{ib1ad})-(\ref{ib5ad}) 
are discretized using the finite difference schemes, quadrature rules and 
discrete $\delta_h$ functions  described in \cite{Peskin02}.

The immersed boundaries are parametrized by the angle $\theta \in [0, 2\pi]$.
We use a mesh $\theta_k = k d\theta$, $k=0,...,{\cal K}$, on them.
To prevent the distances between mesh points which form the immersed boundaries becoming too large as they grow, we increase the number of points in each of them at a certain rate, adding single points (in the case of round shapes) or opposite couples in the lateral walls (in the case of elongated shapes), at the sites where the distance between two neighboring mesh points is larger. This deserves further explanation, since it leads to work with a non uniform angle mesh and with angle dependent elastic moduli, which change as points are added.
Given a mesh $\theta_k$ for a boundary $\mathbf X_j$, with steps 
$d \theta_k = \theta_{k} - \theta_{k-1}$, 
$k=1,...,{\cal K}$, we include a new point between sites $i-1$ and $i$ as follows:
\begin{itemize}
\item Set $d \theta_i =  d \theta_i/2$, $d \theta_{i+1}
=  d \theta_i/2$, and $d \theta_{i+m} = d \theta_{i+m-1},$ $1<m<{\cal K}-i+1$.
\item Set $\theta_i =  \theta_{i-1} + d \theta_i$, $ \theta_{i+1}
 =  \theta_{i} + d \theta_{i+1}$, and $\theta_{i+m} = \theta_{i+m-1},$ $1<m<{\cal K}-i+1$.
\item Set $\mathbf X_j(\theta_i) =  {\mathbf X_j(\theta_{i-1})  +\mathbf X_j(\theta_i) \over 2} $, 
and $\mathbf X_j(\theta_{i+m}) = \mathbf X_j(\theta_{i+m-1}),$ $0<m<{\cal K}-i+1$.
\item Set $K_j(\theta_i)=  2 K_j(\theta_i)$, $K_j(\theta_{i+1})=  2K_j(\theta_i)$, and 
$K_j(\theta_{i+m}) = K_j(\theta_{i+m-1}),$  $1<m<{\cal K}-i+1$, to prevent the reduction in the 
angle from changing the continuum limits.
\item Set ${\cal K} = {\cal K} +1.$
\end{itemize}
Additionally, we need rules for killing cells and dividing cells, which we detail next.

\tikzstyle{block} = [rectangle, draw, fill=blue!20, 
    text width=10em, text centered, rounded corners, minimum height=2em]
\tikzstyle{all} = [diamond, draw, fill=blue!20, 
    text width=6em, text centered, rounded corners, minimum height=1em]
\tikzstyle{decision} = [cloud, draw, fill=blue!20, 
    text width=9em, text centered, rounded corners, minimum height=1em]
\tikzstyle{decision2} = [cloud, draw, fill=blue!20, 
    text width=7em, text centered, rounded corners, minimum height=1em]
\tikzstyle{extreme} = [cloud, draw, fill=red!20, 
    text width=10em, text centered, rounded corners, minimum height=1em]
    
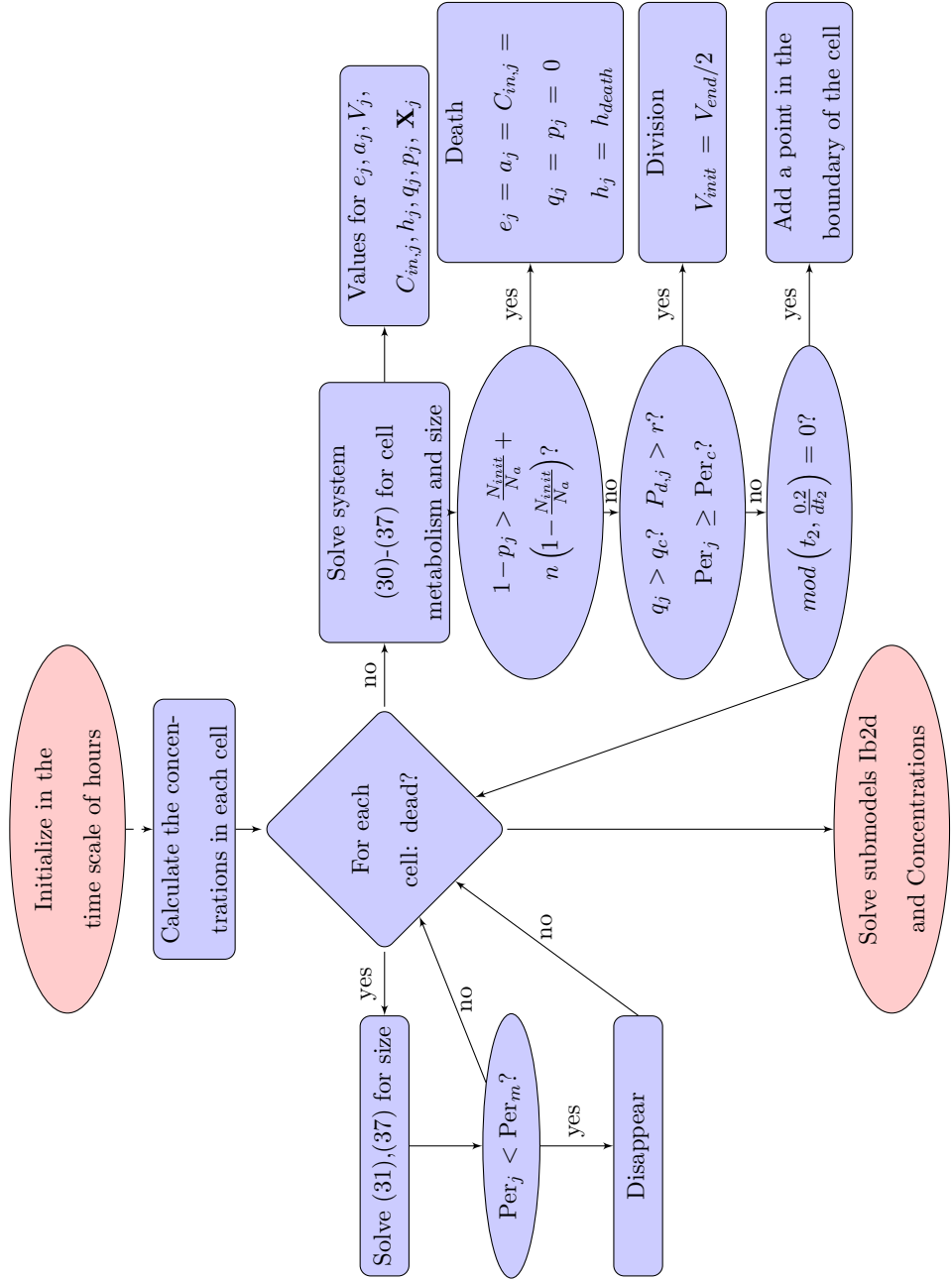
\begin{figure}[h!] \centering
\begin{tikzpicture}[node distance = 2.5cm, auto, rotate=90, transform shape, scale=0.85]
    \node [extreme] (init) {Initialize in the time scale of hours};
    \node [block, below of=init, node distance=2cm] (concen) {Calculate the concentrations in each cell};
    \node [all, below of=concen, node distance=3cm] (isdeath) {For each cell: dead?};
    \node [block, right of=isdeath, node distance=5cm] (system) {Solve system (\ref{deb1ad})-(\ref{deb8ad}) for cell metabolism and size};
    \node [block, right of=system, node distance=5cm] (Vables) {Values for $e_j, a_j, V_j,$ $C_{in,j}, h_j, q_j, p_j,$ $\mathbf X_j$};
     \node [block, left of=isdeath, node distance=5cm] (system2) {Solve (\ref{deb2ad}),(\ref{deb8ad}) for size};
    \node [decision, below of=system, node distance=2.3cm] (DecideDeath) {$1\!-\!p_j \!>\!
    {N_{init} \over N_a} \!+\! n \left(\! 1\!-\!{N_{init} \over N_a} \!\right)?$};
    \node [block, right of=DecideDeath, node distance=6cm] (Death) {Death \\ $e_j=a_j=C_{in,j}=q_j=p_j=0$ \\ $h_j=h_{death}$ };
    \node [decision, below of=DecideDeath, node distance=2.4cm] (DecideDivision) {$q_j> q_c$? \;$P_{d,j}>r$?    ${\rm Per}_j\geq {\rm Per}_c$?};
    \node [block, right of=DecideDivision, node distance=6cm] (Division) {Division\\ $V_{init}=V_{end}/2$};
    \node [decision, below of=DecideDivision, node distance=2cm] (DecideExtention) {$mod\left(t_2,{0.2 \over dt_2}\right)=0?$};
    \node [block, right of=DecideExtention, node distance=6cm] (Extention) {Add a point in the boundary of the cell};
    \node [extreme, below of=isdeath, node distance=8cm] (Ib2d) {Solve submodels  Ib2d and Concentrations};
    \node [decision2, below of=system2, node distance=2cm] (DecideDisappear) {${\rm Per}_j<{\rm Per}_m?$};
    \node [block, below of=DecideDisappear, node distance=2cm] (Disappear) {Disappear};
    \path [line,dashed] (init) -- (concen);
    \path [line] (concen) -- (isdeath);
    \path [line] (isdeath) -- node [midway, above] {no} (system);
    \path [line] (isdeath) -- node [midway, above] {yes} (system2);
    \path [line] (system2) -- (DecideDisappear);
    \path [line] (DecideDisappear) -- node [midway, right] {yes} (Disappear);
    \path [line] (DecideDisappear) -- node [midway, below] {no} (isdeath);
    \path [line] (Disappear.east) -- node [midway, right] {no} (isdeath);
    \path [line] (system) -- (Vables);
    \path [line] (system) -- (DecideDeath);
    \path [line] (DecideDeath) -- node [midway, above] {yes} (Death);
    \path [line] (DecideDeath) -- node [midway, right] {no} (DecideDivision);
    \path [line] (DecideDivision) -- node [midway, above] {yes} (Division);
    \path [line] (DecideDivision) -- node [midway, right] {no}(DecideExtention);
    \path [line] (DecideExtention) -- node [midway, above] {yes} (Extention);
    \path [line] (DecideExtention.west) -- node {}(isdeath);
    \path [line] (isdeath) -- (Ib2d);
\end{tikzpicture}  
\caption{Flowchart for cell evolution in the time scale of hours.}
\label{fig4} 
\end{figure}

\begin{figure}[h!] \centering
\centering
\includegraphics[width=14cm]{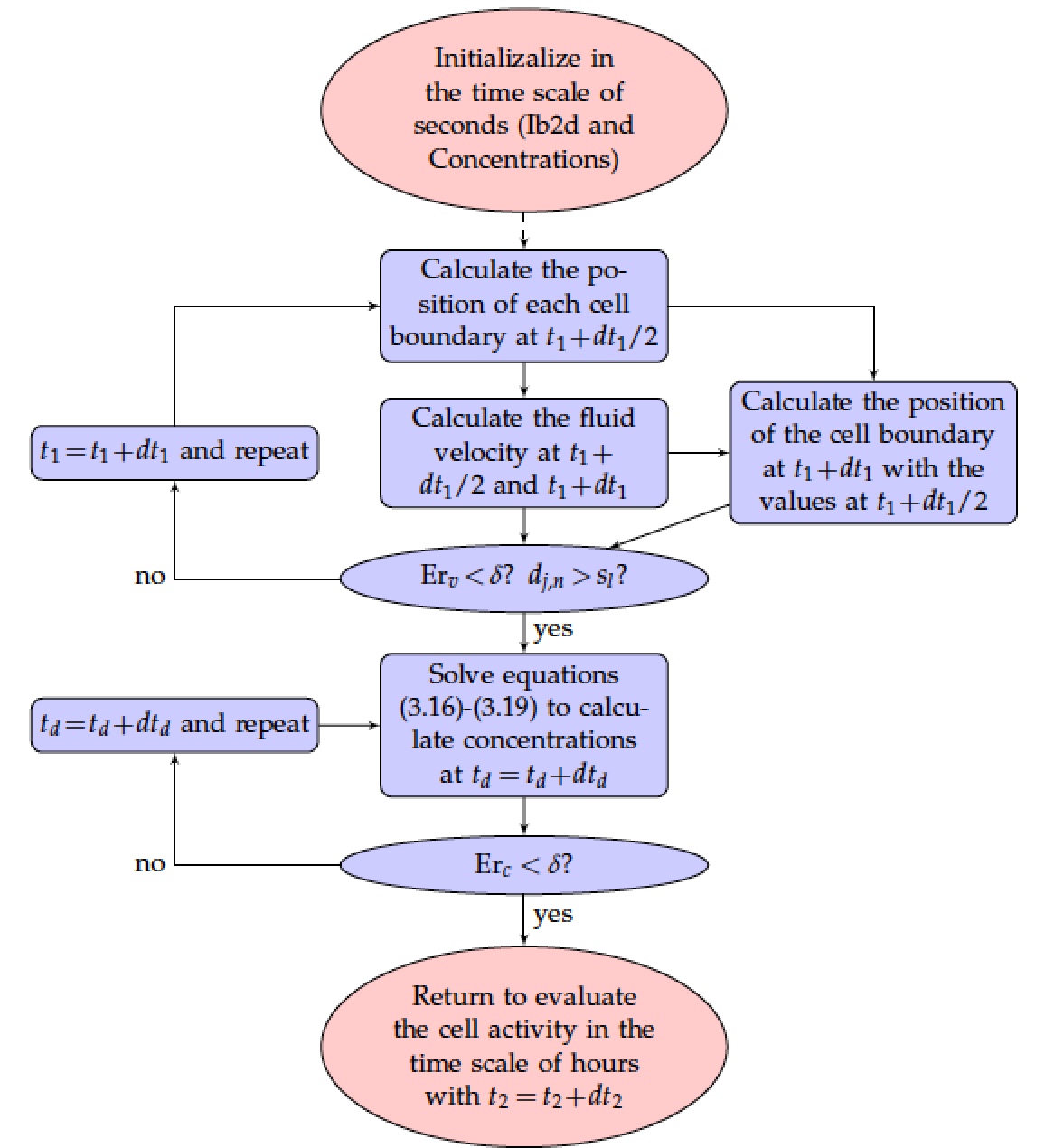}
\caption{Flowchart for the submodels governing IB and chemical processes.
}
\label{fig8}
\end{figure}

\subsection{Rules for division and death}
\label{sec:killdivide}

\begin{figure}[h!]
\centering
\includegraphics[width=14cm]{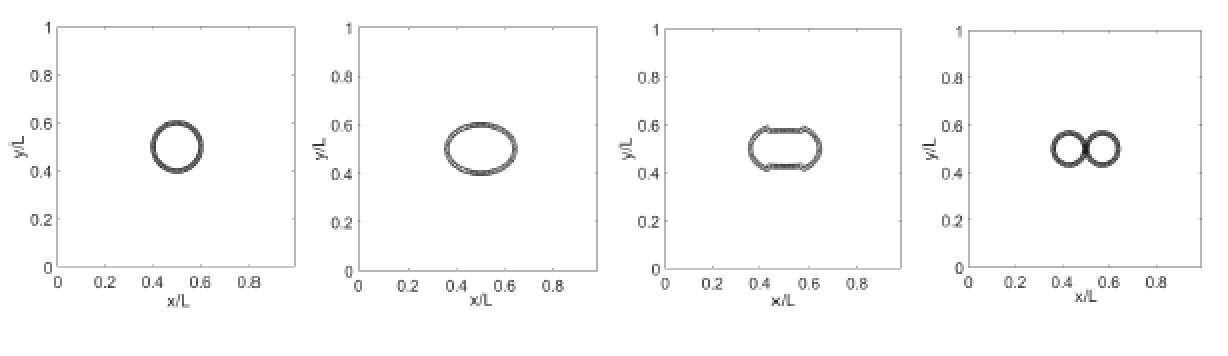}
\caption{Snapshots of the division of a spherical bacterium.}
\label{fig5}
\end{figure}

\begin{figure}[h!]
\centering
\includegraphics[width=14cm]{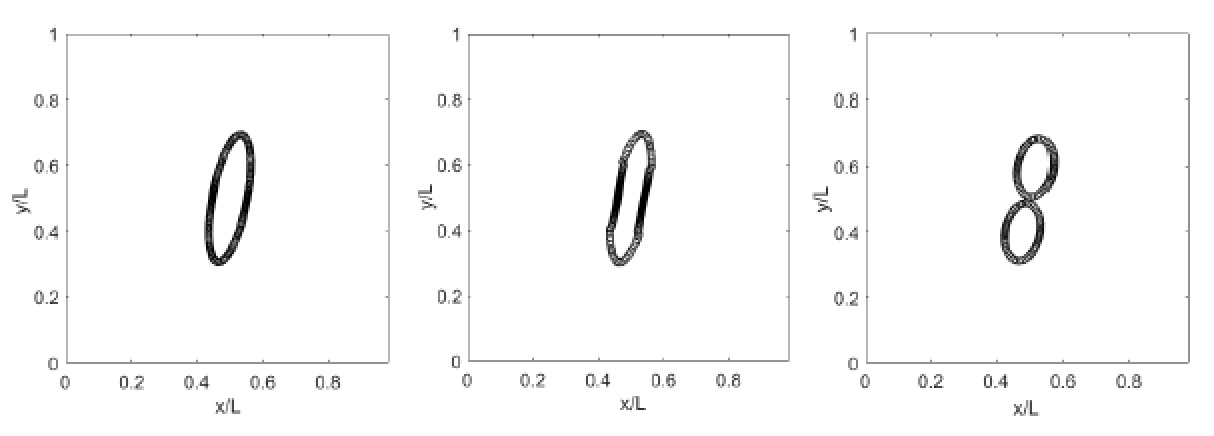}
\caption{Snapshots of the division of a rod-like bacterium. }
\label{fig6}
\end{figure}

Once the size of a bacterium $\mathbf X_j$ surpasses a critical perimeter, the cell divides with probability $P_{d,j}={S_j \over S_j+1},$ $S_j$ being the averaged value of the limiting  concentration at the cell location, provided their aging acceleration $q_j$ is larger than a critical value $q_c$  (a way to indicate age, not to kill newborn cells). More precisely, for each cell boundary $\mathbf X_j$:
\begin{itemize}
\item We check whether $q_j> q_c=10^{-8}$.
\item We check whether its length ${\rm Per}_j$ is larger than a critical perimeter ${\rm Per}_c=1.4 \, {\rm Per}_{init,max}$ for sphera and ${\rm Per}_c=1.5 \, {\rm Per}_{init,max}$ for rod-like bacteria, where ${\rm Per}_{init,max}$ is the maximum perimeter in the initialization step.
\item We generate a random number $r\in(0,1)$ and check whether $P_{d,j}>r.$
\end{itemize}
Figures \ref{fig5} and \ref{fig6} illustrate the division process for spherical and rod-like bacteria.
Division is completed in a few steps: the cell elongates and then splits conserving area.  For spherical bacteria, if $V_{init}= \pi R_{init}^2$ is the volume before  division, we have radius $R_{end}={R_{init} \over \sqrt{2}}$ for the two daughters. For rod-like bacteria, with initial volume 
$V_{init}= \pi b a_{init}$, being $b$ the smallest semi-axis, we have  $a_{end}={a_{init} \over 2}$ for the two daughters, because $b$ is constant. We reset all the cell variables to their initial values after division, see Section  \ref{sec:initialization}.

\begin{figure} [h!] \centering
\includegraphics[width=14cm]{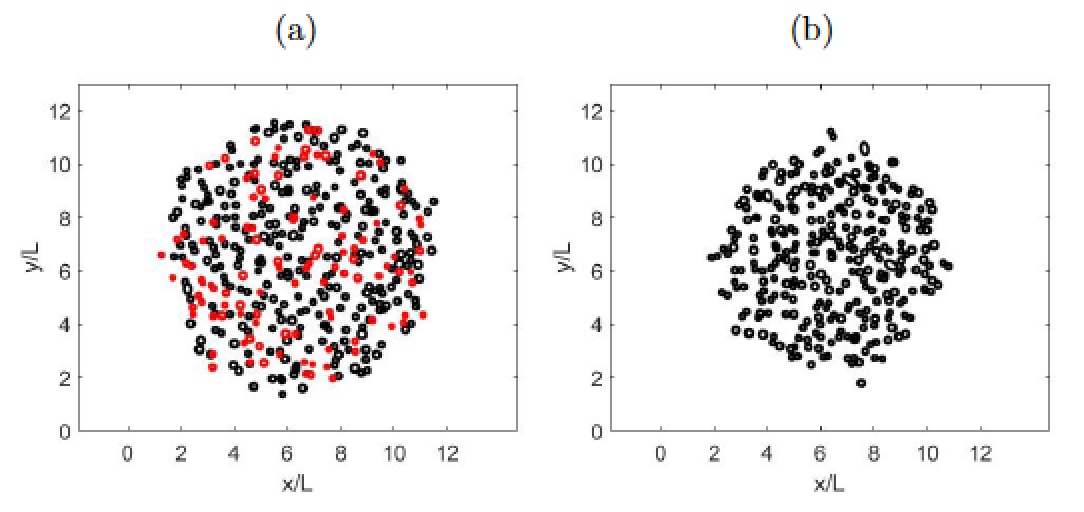} 
\caption{Snapshots of the evolution of a circular biofilm formed by $100$ cells
under the same conditions but different death treatment:
(a) Shrinking dead cells (represented in red). 
We have $292$ alive cells and $108$ dead ones.
(b) Erasing dead cells. 
We have $294$ alive cells and have erased $86$.  
}
\label{fig7}
\end{figure}

Similarly, the cell $\mathbf X_j$ dies with probability defined by $p_j$, $j=1,...N.$ 
We kill $\mathbf X_j$ when $1-p_j>{N_{init} \over N_a} + r\left(1-{N_{init} \over N_a}\right)$, where  $N_a$ is the current number of bacteria, $N_{init}$ the initial number of bacteria and $r \in (0,1)$ a random number.
When a bacterium dies we have two options: 1) erase the cell immediately, 2) keep it and solve only equations (\ref{deb2ad}) for the volume, so that it shrinks slowly due to reabsorption, see Figure \ref{fig7}. The latter option may produces a more realistic evolution in some cases, to account for necrotic regions which otherwise would be erased.
We solve the whole set of equations (\ref{deb1ad})-(\ref{deb8ad}) for the living cells, but only Eq. (\ref{deb2ad}) for the dead cell, fixing $h=h_{death}.$
For spheras, when the dead cell's perimeter is below a minimum threshold ${\rm Per}_m=\pi dx$, $dx$ being the spatial discretization step, the cell disappears. 
For rod-like bacteria we take ${\rm Per}_m=2\pi b$, being $b$ the shortest semi-axis.
The parameter $h_{death}$ governs the speed of the perimeter decrease.
We choose to increase $h_{death}$ with the number of alive cells surrounding the dead one, since it represents reabsorption. More precisely, we set $h_{death,j}=h_{death,j} + dt {\rm Nc}_{C_i<Rd}$, where ${\rm Nc}_{C_i<Rd}$ is the number of cells whose center lies at a distance smaller than $Rd={3/L}$ for cell $j$ and $dt=dt_2.$  

\begin{figure}[h!] \centering
\includegraphics[width=14cm]{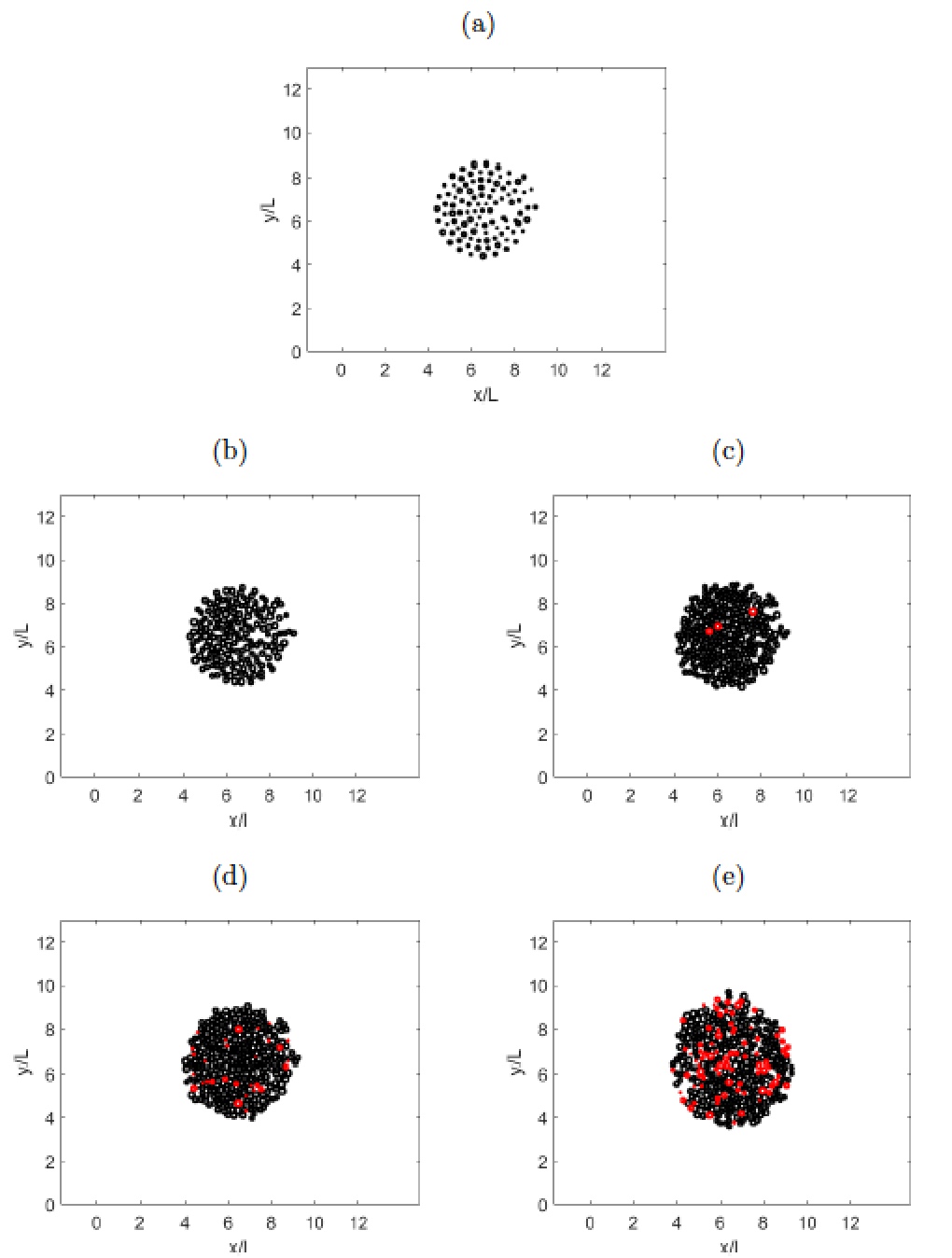} 
\caption{(a) Initial arrangement. Evolution at (b) $t=9$ h, (c) $t=12$ h, (d) $t=14$ h, (e)  $t=20$ h. The simulation starts with $100$ cells and ends up with $286$ cells alive, $81$ dead (red) and  $179$ already erased, see Video3.}
\label{fig9}
\end{figure}

\subsection{Initialization and boundary conditions}
\label{sec:initialization}

A typical geometry initialization is represented in Figure \ref{fig9}(a).
We define $N$ non overlapping immersed boundaries (either spheres or rods) in the region $13L \times 13L$ for sphera and $17L \times 17L$ for rod-like bacteria, located inside a circle of a given radius. The centers, dimensions, axis orientation (when required), and number of points forming the boundaries, vary randomly about given values.  Next, 
\begin{itemize}
\item We create the cubic mesh of step $dx$ in that region to discretize the fluid and the reaction-diffusion equations.
\item We set the initial velocity $\mathbf u$ equal to zero everywhere and periodic 
boundary conditions for the fluid velocity.
\item A reference value $S_0=10$ is fixed as initial and Dirichlet boundary condition for the concentration at the borders of the computational region. 
\item We set $C_e(0) =0$ and $\varepsilon(0)=0$ everywhere and enforce zero 
Neumann boundary conditions for them.
\item For the first simulations, we set  $C_{out}(0)=0$ everywhere and 
enforce zero Dirichlet boundary conditions. 
Once the biofilm seed has evolved for some time, we switch to a Dirichlet boundary condition $C_{out}= 3,\, 7,\, 30$ on the borders of the computational
region. As initial condition for $C_{out}$ we use the profile obtained by relaxation of (\ref{c3ad}) with the boundary condition and without the convective term.
\item For $j=1,...,N$ we set $V_j(0)$ equal to the initial dimensionless areas, 
$e_j(0)= {S_j(x_j,0) \over S_j(x_j,0) +1}$, $x_j$ being the center of
cell $j$, $V_{e,j}(0)=0$, $q_j(0)=0$, $h_j(0)=0.6$, $p_j(0)=1$, $a_j(0)=0$, and
$C_{in,j}(0)=0$. When we divide a cell, they start with the same initial conditions, 
except $C_{in,j}$  in the presence of a toxicant, which divides a random percentage to one and the opposite to the other.
\end{itemize}

\begin{figure}[h!] \centering
\includegraphics[width=14cm]{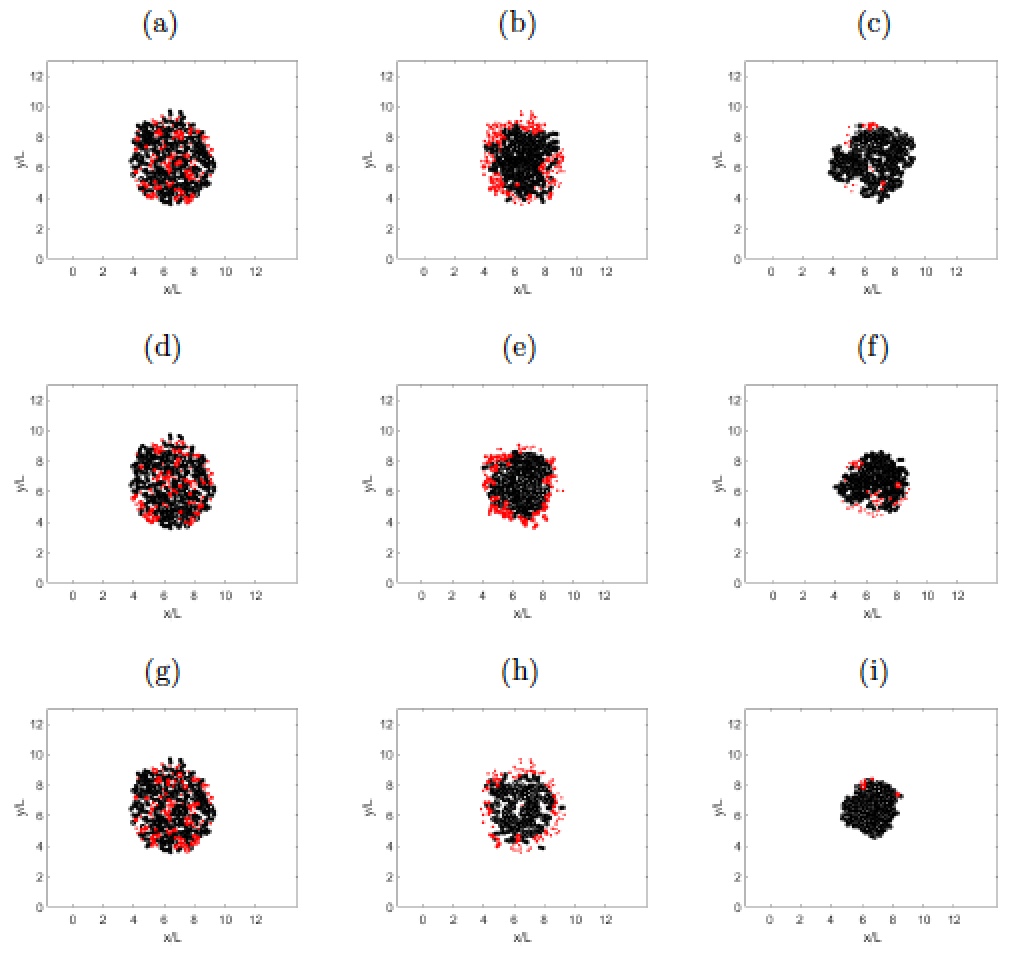}
\caption{Evolution of the final configuration reached in Fig. \ref{fig9} under the action of antibiotics.
Snapshots for $C_{out}= 3$  at (a) $t=1$ h, (b) $t=6$ h, and (c) $t=12$ h. The simulation ends with $260$ cells alive, $17$ dead (red) and $419$ already erased, see Video3a. 
Snapshots for $C_{out}= 7$ at  (d) $t=1$ h, (e) $t=5$ h, and (f) $t=12$ h. The simulation end with $213$ cells alive, $38$ dead (red) and $390$ already erased, see Video3b. 
Snapshots for $C_{out}= 30$ at (g)  $t=1$ h, (h)  $t=3.5$ h, and (i) $t=10$ h. The simulation ends with $162$ cells alive, $6$ dead (red) and $354$ already erased, see Video3c.
If we do not erase them, we have a necrotic outer layer of increasing thickness.}
\label{fig10}
\end{figure}

\begin{figure}[h!] \centering
\includegraphics[width=14cm]{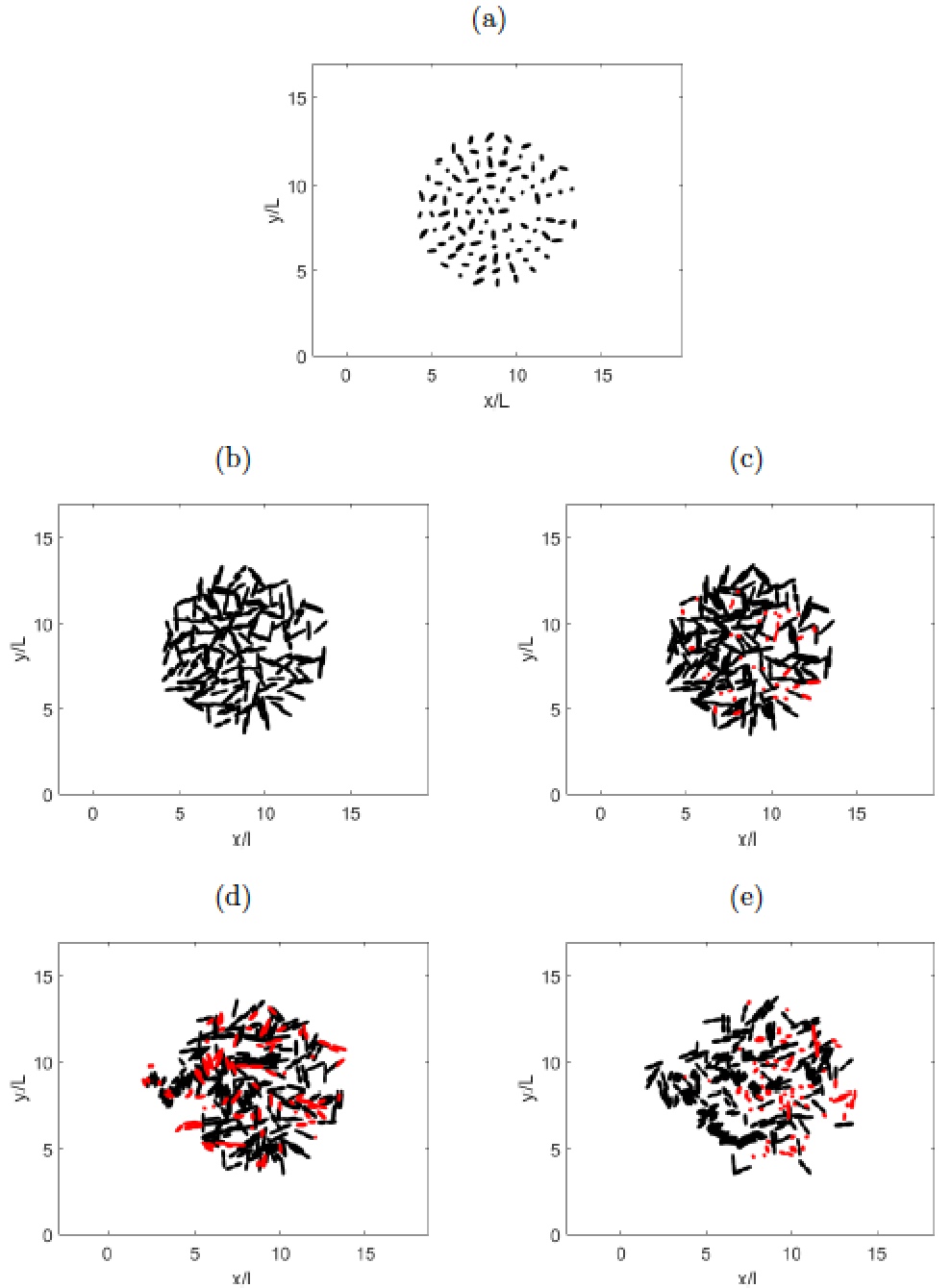}
\caption{(a) Initialization. Evolution at (b) $t=9.5$ h, (c) $t=12$ h, (d) $t=14$ h 
and (e) $t=20$ h, without antibiotics. We started with $100$ bacteria, and ended 
with $289$ alive, $68$ dead (red) and $267$ disappeared, see Video4.
}
\label{fig11}
\end{figure}

\begin{figure}[h!] \centering
\includegraphics[width=14cm]{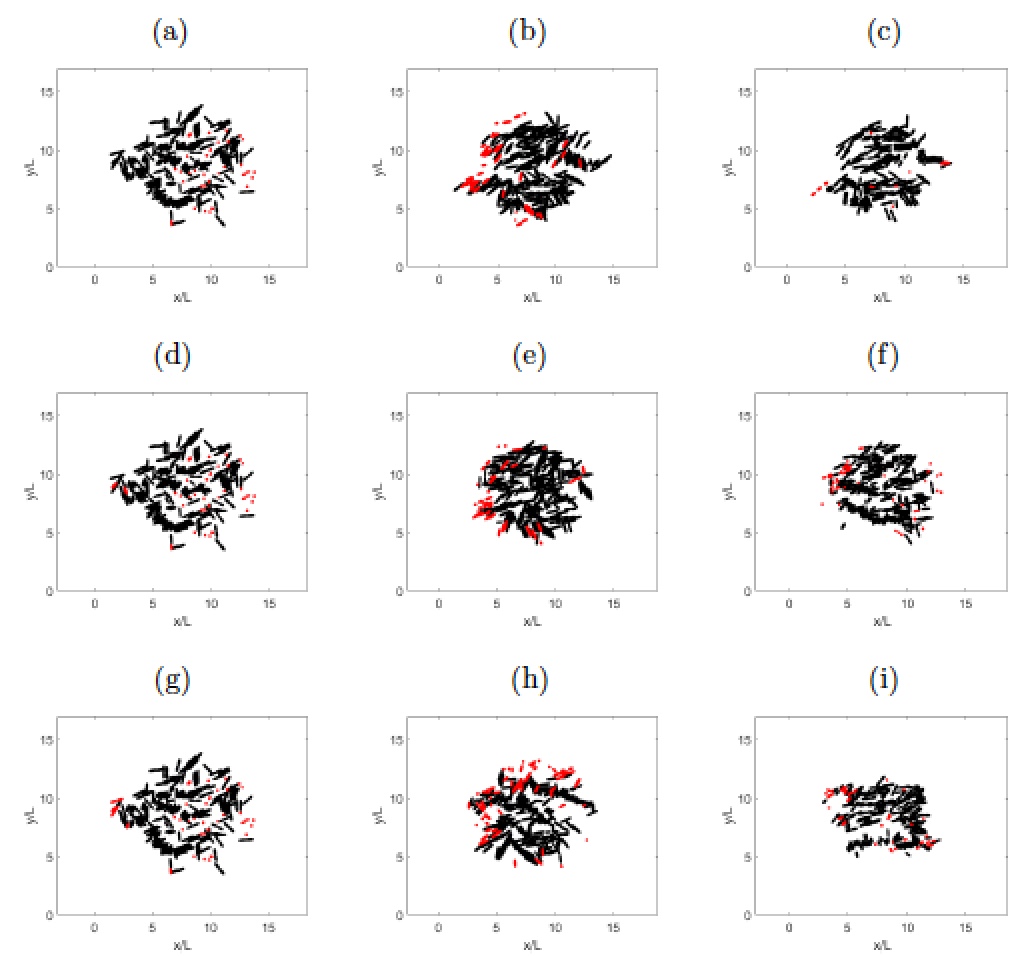}
\caption{Evolution of the final configuration reached in Fig. \ref{fig11} under the action of antibiotics.
Snapshots for $C_{out}= 3$ at (a) $t=1$ h, (b) $t=6.5$ h, and (c) $t=10$ h with $284$ alive, $14$ dead (red) and $358$ erased, see Video4a.
Snapshots for  $C_{out}= 7$ at (d) $t=1$ h, (e) $t=7.5$ h, and (f) $t=10$ h with $278$ alive, $40$ dead (red) and $340$ erased, see Video4b. 
Snapshots for $C_{out}= 30$ at (g)  $t=1$ h, (h)  $t=3$ h, and (i) $t=10$ h with $246$ alive, $34$ dead (red) and $346$ erased, see Video4c. 
}
\label{fig12}
\end{figure}

Figures \ref{fig9}-\ref{fig10} show a few snapshots of the evolution of a circular biofilm formed by spherical bacteria, without antibiotic and with antibiotics, respectively,  see also Videos 3, 3a, 3b, 3c.
The action of antibiotics would vary depending on parameters we have fixed, such as the toxicity, and the parameters governing the flux inside and outside the cells. We see that as the antibiotic presence is increased, growth slows down, less cells remain, and an outer necrotic region appears, that finally dissolves in the surrounding fluid and is absorbed by the remaining cells.
The dynamics of dead cells depends on the governing parameters we choose to govern the reabsorption process. Figures \ref{fig11}-\ref{fig12} illustrate the 
evolution for rod-like bacteria, see also Videos 4, 4a, 4b, 4c.

As said earlier, we use a specific discretization of the Inmersed Boundary model, solving (\ref{ib1ad})-(\ref{ib6ad}) by Fourier transforms \cite{Peskin95,Peskin02}. We use the time $t_1$ as an artificial time until the system relaxes to a stationary state, with step $d t_1 = 10^{-6}$. When the relative errors of the fluid-IB variables ${\rm Er}_v$ fall below a tolerance  $\delta$, we use the time $t_d$ as an artificial time until the concentration system relaxes to a stationary state with a step $d t_d = 10^{-9}$ for spheres and $d t_d = 10^{-11}$ for rods, due to the convection factor $T_d/T=10^{6}$. When the relative errors  ${\rm Er}_c$ fall below a tolerance $\delta$, we stop. We set $\delta =10^{-3}$.  We also demand that the cells remain at a certain distance $s_l,$ in these tests we have set  $s_l=0.$

\section{Computational model in the presence of barriers} 
\label{sec:barrier}

\begin{figure} [h!]
\centering
\includegraphics[width=14cm]{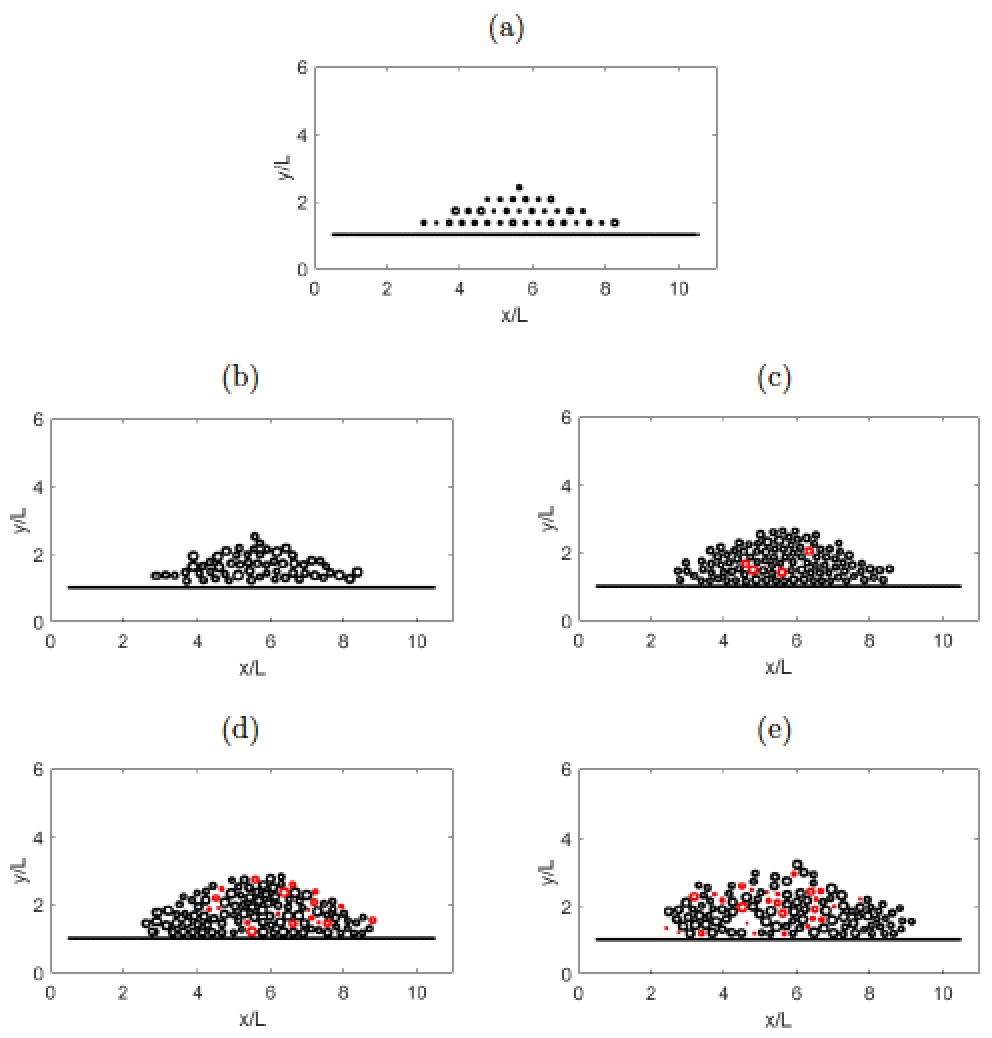} 
\caption{(a) Initial arrangement. Evolution at (b) $t=9$ h, (c) $t=12$ h, (d) $t=15$, (e)  $t=20$ h.
The simulation starts with with 34 cells and ends with  $97$ alive cells, $27$ dead cells (red) and $58$ cells already erased, see Video5. 
}
\label{fig13}
\end{figure}

As mentioned earlier, we are interested in two kinds of two dimensional reductions.
So far, we have considered the horizontal spread of a two dimensional cluster. We focus here  on the arrangement depicted in Figure \ref{fig2}: a biofilm slice expanding on a surface. 
The model equations remain the same as in Sections \ref{sec:nond} and \ref{sec:computational}.
The main change concerns the geometry: we introduce a boundary orthogonal
to the biofilm slice representing the interface on which it grows.
We place  bacteria on a semi-circle on top of it, see Figure \ref{fig13}(a).
We will exploit the strategy developed in Section \ref{sec:computational}, including
additional equations for the horizontal barrier. We impose on it the same equations as for the cell boundaries,  without the growth force, and without interaction force (bacteria do not move the barrier). On the other hand, cells do notice the presence of the barrier and the corresponding interaction is included for them.
Moreover, in equation (\ref{f1b}), in front of the integral, we add a factor 
$0.001$ to account for higher density of the barrier and almost negligible 
barrier mobility due to fluid.

The main variations arise when working with rod-like bacteria. We set $dt_d=10^{-10}$. In this case, forces can generate a moment that rotate bacteria. This force creates a torque, $\mathbf{M}_f$, that then varies the angular momentum $\mathbf{L}$, and knowing the moment of inertia $\mathbf{L}=\mathbf{I}\mathbf{w}$, we obtain the angular velocity $\mathbf{w}$,  $\mathbf{I}$ being the body's inertia tensor.
\begin{eqnarray}
\mathbf{M}_f = \mathbf{X} \times \mathbf{F}_i, \quad {d\mathbf{w} \over dt} = \mathbf{I}^{-1}\mathbf{M}_f. \label{f2ab}
\end{eqnarray}
In two-dimensions, directions of $\mathbf{M}_f$ and $\mathbf{w}$ are perpendicular to the plane. Thus, we only need the moment of inertia of the axis perpendicular to the plane, which is $I={1 \over 4}M(a^2 + b^2)$ for elliptical shapes, where $a$ is the long semi-axis, and $b$ the short one. $M$ is the mass of the bacteria, $M=\rho_x V$, where $\rho_x$ is bacterial density and $V$ its volume. In two-dimensions, they become surface density and area. In this way, we can add in Eq. (\ref{ib3}) the following expression
\begin{eqnarray}
{\partial \mathbf X \over \partial t} = \mathbf{w} \times \mathbf X. 
\label{f2ac}
\end{eqnarray}

When we nondimensionalize, we need to include in the right hand side of Eq. (\ref{ib3ad}) the term $\mathbf{w} \times \mathbf{x}$ with
\begin{eqnarray}
 {d\mathbf{w} \over dt} = C_f \mathbf{I}_0^{-1}\mathbf{M}_f, \quad
 \mathbf{M}_f = \mathbf{x} \times \mathbf{F}_i, \label{f2ab}
\end{eqnarray}
where all terms are dimensionless, and $C_f=T^2{E_s L^2 \over \rho_{x,s} L^2}={150 \over 47}10^{-6}$ is a dimensionless number, calculated using ${E_s \over \rho_{x,s}} = {E \over \rho_{x}}$. Moreover, $\mathbf{I}_0={1 \over 4}M_0(a_0^2 + b_0^2)$, where $M_0=V$, $V=\pi a_0b_0$ is dimensionless  bacterial area and $a_0=a/L$, $b_0=b/L$.

A new feature we wish to represent in this new set-up is the observation that fluid flows upwards through the horizontal barrier because the bacterial biofilm seed swells. We are representing the threads keeping together bacteria in the biofilm as interaction forces keeping bacteria at a distance. When the biofilm swells, those threads swell and elongate too. We model this fact changing the minimum distance between bacteria in the biofilm.

\begin{figure}[h!] \centering
\includegraphics[width=14cm]{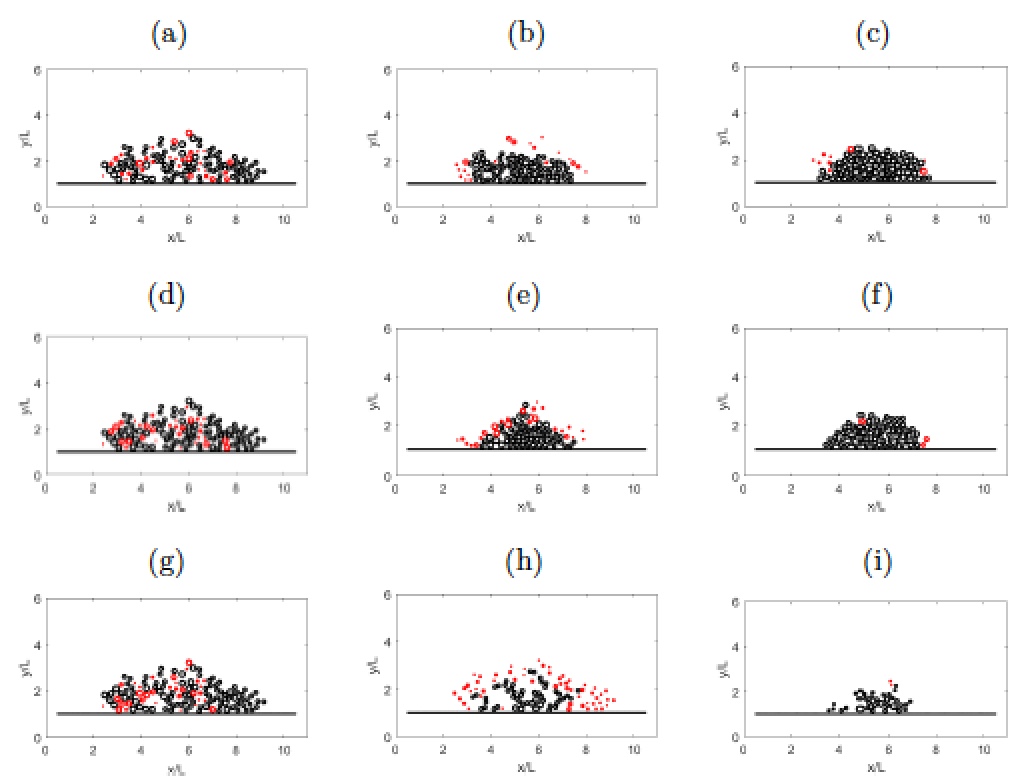}
\caption{ Evolution of the final configuration reached in Fig. \ref{fig13} under the action of antibiotics. 
Snapshots for $C_{out}= 3$ at (a) $t=1$ h, (a) $t=7.5$ h, and (c) $t=10$ h.
The simulation ends with $77$ cells alive, $7$ dead (red) and $120$ erased, see Video5a. 
Snapshots for $C_{out}= 7$ at (d) $t=1$ h, (e) $t=7$ h, and (f) $t=10$ h.
The simulation ends with $69$ cells alive, $3$ dead (red) and $125$ erased, see Video5b. 
Snapshots for $C_{out}= 30$ at (g)  $t=1$ h, (h)  $t=3.5$ h, and (i) $t=10$ h.
The simulation ends with with $34$ cells alive, $2$ dead (red) and $116$ erased, see Video5c.
}
\label{fig14}
\end{figure}

\begin{figure} [!h]
\centering
\includegraphics[width=14cm]{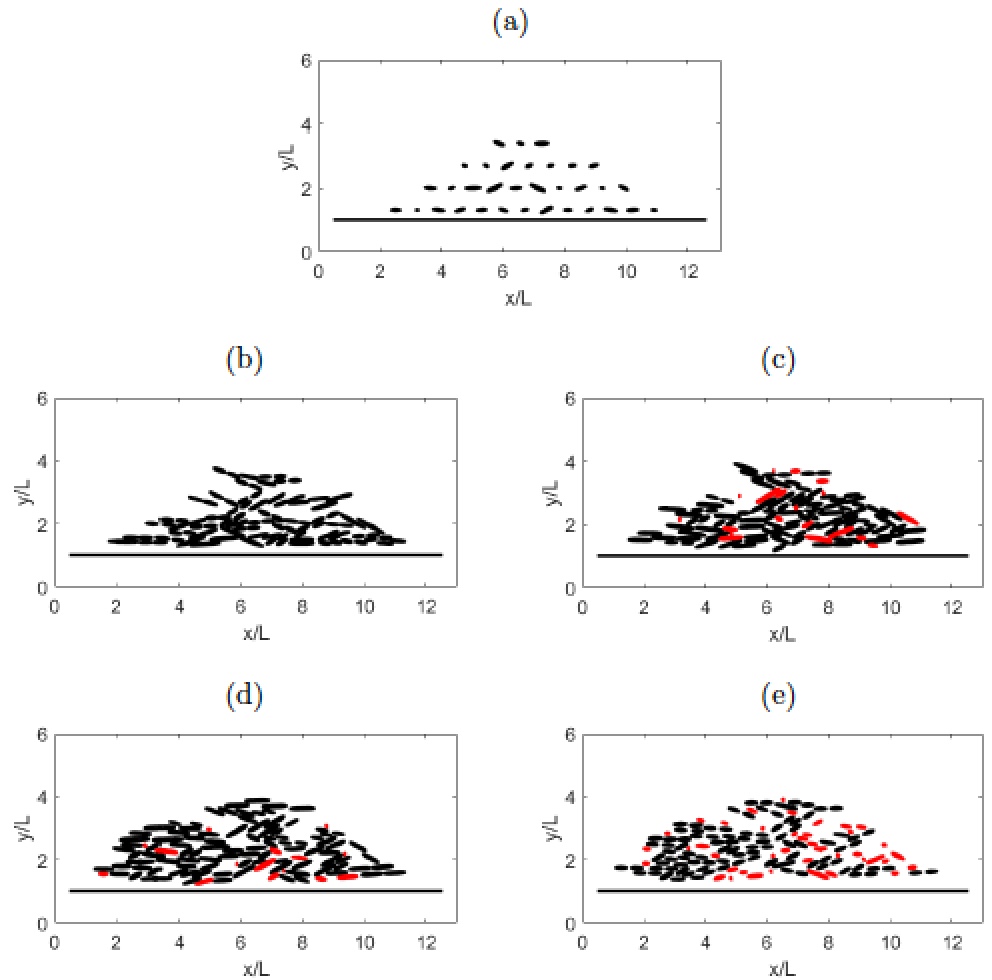}
\caption{(a) Initial arrangement. Evolution at (b) $t=10$ h, (c) $t=14$ h, (d) $t=18$ h, (e)  $t=20$ h.
The simulation starts with  $33$ cells and ends with $96$ cells alive, $41$ dead and $75$ erased, see Video6.
}
\label{fig15}
\end{figure}

\begin{figure}[!h] \centering
\includegraphics[width=14cm]{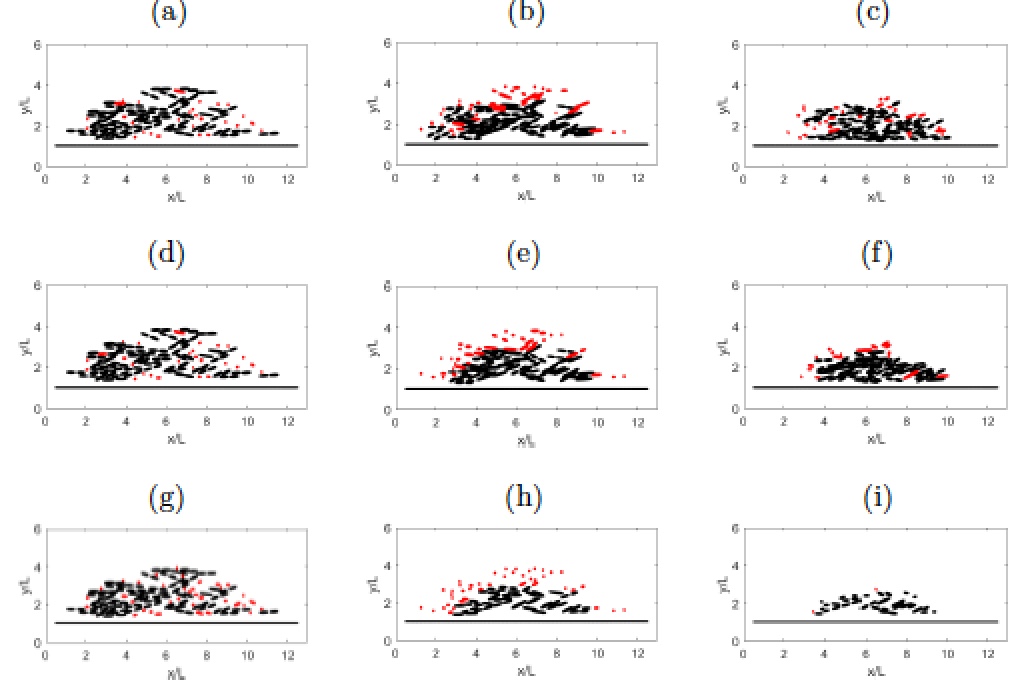}
\caption{Evolution of the final configuration reached in Figure \ref{fig15} under the action of
antibiotics. Snapshots for $C_{out}= 3$ at (a) $t=1$ h, (a) $t=4$ h, and (c) $t=10$ h.
The simulation ends with  $84$ cells alive, $29$ dead (red) and $108$ erased, see Video6a. 
Snapshots for $C_{out}= 7$ at (d) $t=1$ h, (e) $t=4$ h, and (f) $t=10$ h with $58$ alive, 
The simulation ends with $17$ cells dead (red) and $115$ erased, see Video6b. 
Snapshots for $C_{out}= 30$ at (g)  $t=1$ h, (h)  $t=3.5$ h, and (i) $t=10$ h.
The simulation ends with $33$ cells alive, $2$ dead (red) and $105$ erased, see Video6c.
}
\label{fig16}
\end{figure}

For spherical bacteria, we modify the repulsive force because it is not the same to push upwards than horizontally without the force of gravity. The force is of lesser magnitude and the repulsion occurs more gradually:
\begin{eqnarray}
\mathbf F_i = \sum_{j=1}^N \mathbf F_{i,j} \delta_j, \quad
\mathbf{F}_{i,j} = \displaystyle\sum_{n=1,n\neq j}^{N-1} \sigma_b \, e^{-{d_{j,n}^2 \over l_{sp}}} \mathbf{n}_{{\rm cm},n,j}, \label{f2b} 
\end{eqnarray}
$\sigma_b$ is the repulsive parameter, and $l_{sp}$ sets the maximum distance, where the cells begin to repel. The latter term changes over time, as swelling causes the strings that separate the cells to grow. We have set
\begin{eqnarray}
l_{sp} = l_{max} {1+\tanh\left(v_s \left(t-T_p\right)\right)\over 2} \quad s_{p2} = \sqrt{-\ln(v_{min})l_{sp}}, \label{f2c}
\end{eqnarray}
where $l_{max} = -{s_{max}^2 \over \ln(v_{min})}$ and $v_s$ is related to the growth of this distance.
It saturates at a certain time, we use an inflection point $T_{p}$, and a certain maximum length $l_{max}$. This value depends on the maximum separation of the cells $s_{max}$ and a minimum variation $v_{min}$. All of this affects the critical distance $s_{p2}$. All cells tend to be more or equal apart.
Removing dimensions, the interaction force is as follows:
\begin{eqnarray}  
\mathbf F_i = \displaystyle\sum_{j=1}^N \displaystyle\sum_{n=1,n\neq j}^{N} \sigma_{b,0} \, e^{-{d_{j,n}^2 \over l_{sp}}} \delta_j \mathbf{n}_{{\rm cm},n,j}, \label{ib7ad}
\end{eqnarray}
where $\sigma_{b}=\sigma_{b,0}LE_s=20E_s$, so $\sigma_{b,0}=20/L$. And $\tilde l_{sp}(t_1)=l_{sp}(t_1)/L^2$. We drop the symbol $\, \tilde{} \,$ for ease of notation. Parameters are collected in Table \ref{table8}.

For rods  there is anisotropy, the vertical direction being different from the horizontal one. We set
\begin{eqnarray}
s_{p2} = t{s_{p,m} \over T_{m}}, \label{f2a}
\end{eqnarray}
where ${s_{p,m} \over T_m}$ is the slope to which the distance with respect to time ascends. 
We do not have to change the force because the interaction in one plane and the other are similar, the only difference being the growth of the distance.  
Removing dimensions
\begin{eqnarray}
s_{p2} = t_2{s_{pm,0} \over T_{pm,0}}. \label{fb1}
\end{eqnarray}
In either case, spheres or rods, we set $s_l=s_{p2}$ in the flowchart.

\begin{table}[h!]
\begin{center} \begin{tabular}{|c|c|c|c|c|c|c|}
 \hline
$s_{max,0}={s_{max} \over L}$ & $l_{m,0}=-{s_{max}^2 \over \ln(v_{min})L^2}$  &  $T_{p,0}={T_p \over \tau d t_2}$ & $v_{min}$ 
 \\ \hline
$0.04$ & $-{1.6 \times 10^{-3} \over \ln(v_{min})}$ & $6.5 \over d t_2$ & ${d t_2 \over 2 d t_1 \lambda_0 \sigma_{b,0}}$ 
 \\ \hline
 $s_{pm,0}={s_{pm} \over L}$ & $T_{pm,0}={T_{pm} \over \tau d t_2}$ & $v_{s,0}=\tau v_s$ & 
 \\ \hline
$0.04$ & $11 \over d t_2$ & $5 \times 10^{-3}$ &
 \\ \hline
\end{tabular} \end{center}
\caption{Additional parameters for the simulations in the presence of an horizontal barrier.}
\label{table8}
\end{table}

In this second geometry, nutrients flow to bacteria through the horizontal immersed boundary  on top of which they grow, whereas toxicants flow from the top. As for the initialization, besides the $N$ immersed boundaries representing bacteria, we include a lower barrier which  does not touch the borders of the computational region. Boundary conditions for concentrations change. We fix Dirichlet boundary conditions for $S$ and $C_e$ on the lower computational border, and on the lateral ones up to the height of the horizontal immersed boundary. Zero Neumann boundaries are imposed on the rest. For $C_{out}$, the situation is reversed. Zero Neumann boundary conditions on the lower part, and Dirichlet on the upper one. 

Figures \ref{fig13} and \ref{fig14} illustrate the evolution in the case of spherical bacteria, with and without antibiotics. Notice the formation of inner gaps or channels in the structure. When antibiotics are added, outer necrotic region finally erased appear too. Figures \ref{fig15} and \ref{fig16} illustrate the evolution in the case of  rod-like bacteria.

\section{Biofilm extinction} 
\label{sec:extinction}

\begin{figure}[!h] \centering
\includegraphics[width=14cm]{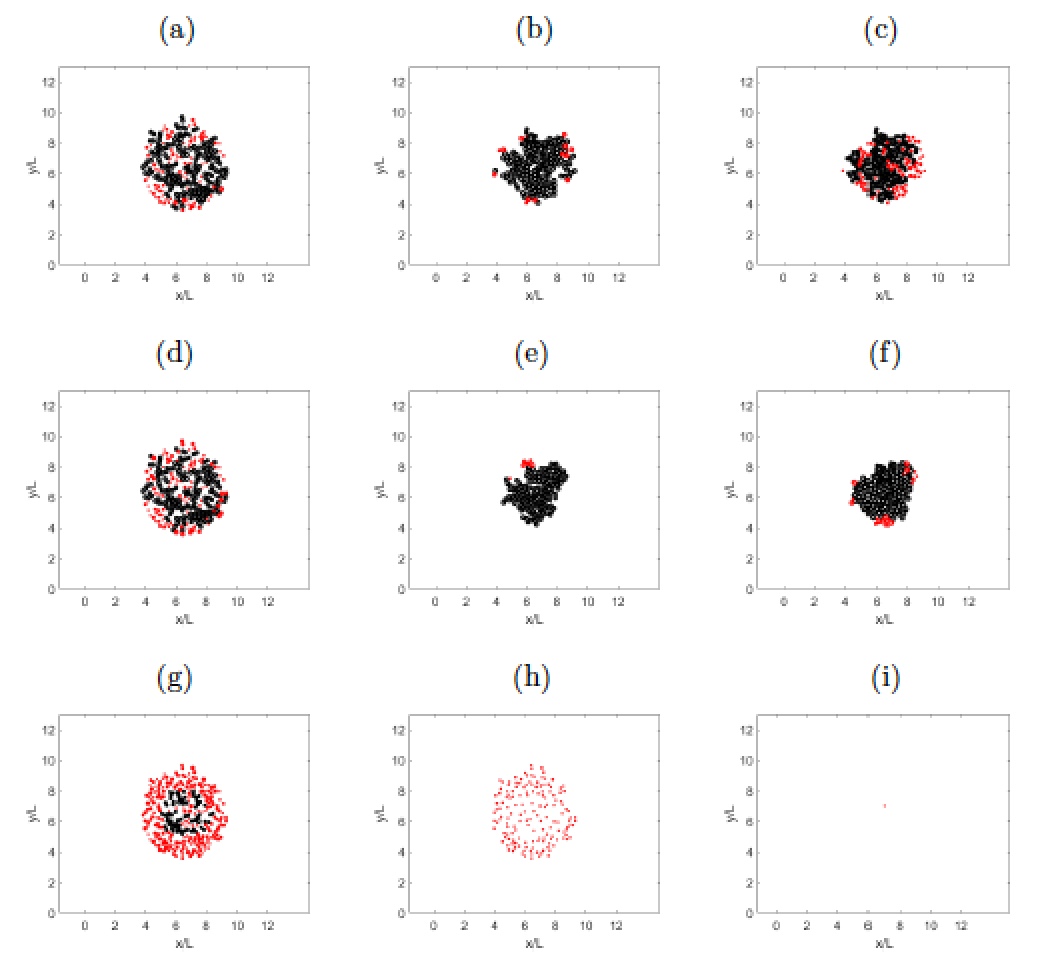}
\caption{Equivalent of snapshots (a)-(c) and (d)-(f) with the modified death criterion:
Snapshots for $C_{out}= 3$ at (a) $t=2.5$ h, (a) $t=8.5$ h, and (c) $t=10$ h.
The simulation ends with  $184$ cells alive, $66$ dead (red) and $320$ erased, see Video7. 
Snapshots for $C_{out}= 7$ at (d) $t=2.5$ h, (e) $t=8.5$ h, and (f) $t=10$ h with $141$ alive, 
The simulation ends with $20$ cells dead (red) and $327$ erased, see Video8. 
Finally, panels (g)-(i) represent the extinction of the same initial configuration with the
modified death criterion and a combination  of two antibiotics with
 $C_{out}= 3$ and variable $k_{tox}$: (g)  $t=2.5$ h, (h)  $t=8.5$ h, and (i) $t=10$ h.
The simulation ends with $0$ cells alive, $1$ dead (red) and $385$ erased, see Video9.
}
\label{fig17}
\end{figure}

In this Section, we consider the possibility of driving a biofilm to extinction by an
adequate combination of antibiotics \cite{Hoiby}. The death criterion we employed
in the previous sections allows the biofilm to grow but it prevents the total number
of bacteria from dropping below the initial value. For decaying biofilms, the
death criterion used in \cite{Deb18} is more adequate: we kill  a cell $\mathbf X_j$ 
when $p_j<r{N_a \over N_{init}}$, being $N_{init}$ the number of bacteria just before administering the antibiotics. In Figure \ref{fig17}, we revisit simulations (a)-(c) and 
(d)-(f) from Figure \ref{fig9} with this new criterion. 
Clinical tests \cite{Hoiby} point out the convenience of combining antibiotics targeting
different types of cells within the biofilm to be able to eradicate them. We consider 
here a cocktail of two antibiotics. One of them targets dormant cells with little energy,
which are located in the inner biofilm core (the antibiotic colistin, for instance). We 
represent that effect using a toxicity coefficient $k_{tox,1,j}$ which decreases wth the 
cell energy. The other one targets cells with high energy, which divide actively, 
and tend to be located in the outer biofilm regions (penicillins, for instance). 
We represent that effect by a toxicity coefficient $k_{tox,2,j}$ which increases with the 
cell energy. More precisely, we have used the following expression
\begin{eqnarray}
k_{tox,1,j} = k_{tox} e^{10(e_m-e_j)},
\quad k_{tox,2,j} = k_{tox} e^{10(e_j-e_m)},  \quad e_m=0.5. \label{fext1}
\end{eqnarray}
We modify the model to include two equations similar to (\ref{c3ad}) for the antibiotic concentration with  toxicity coefficients (\ref{fext1}) and the corresponding two equations
(\ref{deb6ad}) for the antibiotic concentration inside the cells. Also, we set
$C_{out}=C_{out,1} + C_{out,2}$ in the definition of  (\ref{deb1ad}) for $\nu'$ 
and replace in eq. (\ref{deb3ad}) the term $k_{tox} C_{in,j}$ by $k_{tox,1,j}C_{in,1,j}
+k_{tox,2,j}C_{in,2,j}$. Revisiting the simulations in Figure \ref{fig9} with these new choices,
we are able to drive the biofilm to extinction, see Figure (\ref{fig17}) (g)-(i).

\section{Conclusions}
\label{sec:conclusions}

Studying the dynamics of cellular aggregates such as bacterial biofilms faces the
challenge of dealing with complicated geometries and interactions. Many approaches have been proposed to that effect, with advantages and disadvantages. Cellular automata allow us to represent many microscopic and macroscopic processes
\cite{Laspidou,poroelastic}, but ignore bacterial shapes and interactions. Individual based  models seem effective for large biofilms growing in flows \cite{Picioreanu_fluids}, but become exceedingly complicated for biofilms spreading on surfaces as the ones we consider here \cite{Picioreanu_surface,Allen}. Immersed boundary methods provide  a very flexible alternative to study mechanical interactions in these complex geometries
\cite{ibm_adhesion,ibm_division,ibm_tumor}.  Here, the immersed boundaries provide the basic geometrical skeleton, while the interaction with the medium is represented by forces governed by a set of equations coupling metabolic and physico-chemical processes. Cell growth, division, and death, is managed through additional rules on the evolution of the discrete boundaries. Unlike previous IB approaches to multicellular tissues, we do not include heuristical sources.
Boundaries move as a result of cellular activity as dictated by a dynamic energy
budget model, letting flow in and out through them.
We have applied this framework to reproduce initial stages of the spread of a biofilm seed formed by a few spherical or rod-like bacteria in two dimensional geometries. Simulating
rod-like bacteria is more expensive computationally. Computing the interactions of rods
requires small steps to let configurations adapt as cells growth and divide avoiding overlaps.
We observe that rod-like bacteria tend to align.
In radial horizontal views, we see how crowded areas trigger the death of scattered bacteria, which are reabsorbed. For vertical slices expanding on an horizontal barrier, we see also gaps created by death bacteria near the barrier. In this case, we have implemented a mechanism to allow water flow  inside the biofilm, so that gaps are filled with fluid and the separation between bacteria increases.
When antibiotics are applied, bacteria located in the borders are first to die, forming small necrotic regions. We have shown that combining antibiotics which target either active or dormant cells within the layered biofilm structure we are able to drive the biofilm to
complete extinction.

The specific results of the simulations depend on the parameters we choose. 
Most of parameters appearing in the model equations are taken from experimental measurements and fittings to population counts for some bacteria. However, there  are a number of parameters in the representation of interaction forces, division and death criteria which are selected to produce adequate results, avoiding artifacts. Whether the whole set of parameters can be fitted to data counts for the time evolution of biofilm seeds of bacteria deserves further research. 
From a practical point of view, it would be important to be able to implement control strategies using the antibiotic supply as control variables to extinguish the whole biofilm seed in finite time. 


\acknowledgments
This research has been partially supported by the FEDER /Ministerio de Ciencia, 
Innovaci\'on y Universidades - Agencia Estatal de Investigación grant No. 
MTM2017-84446-C2-1-R  and by the Ministerio de Ciencia, Innovaci\'on y Universidades  "Salvador de Madariaga" grant PRX18/00112 (AC). A.C. thanks R.E. Caflisch for hospitality during a sabbatical stay at the Courant Institute, NYU, and C.S. Peskin for nice discussions and useful suggestions.


\begin{thebibliography}{9}
\bibitem[$\dag$]{rafael:email} {E-address \tt  rafael09@ucm.es}.
\bibitem[*]{carpio:email} {E-address \tt ana\_carpio@mat.ucm.es}.
Author to whom all correspondence should be addressed.

\bibitem{biofilm} H.C. Flemming, J. Wingender, The biofilm matrix, Nat. Rev. Microbiol. 8, 623--633, 2010.
\bibitem{Hoiby} N. H{\o}iby, T. Bjarnsholt, M. Givskov, S. Molin, O. Ciofu, Antibiotic resistance of bacterial biofilms, Int J Antimicrob Agents 35, 322--32, 2010.
\bibitem{streamers} K. Drescher, Y. Shen, B.L. Bassler, H.A. Stone, Biofilm streamers cause  catastrophic disruption of flow with consequences for environmental and medical systems, Proc. Natl. Acad. Sci. USA 110, 4345--4350,  2013.
\bibitem{Laspidou} C. S. Laspidou, L. A. Spyrou, N. Aravas,  B. E. Rittmann,
Material modeling of biofilm mechanical properties, Math. Biosci. 251, 11-15, 2014.
\bibitem{Picioreanu_fluids} L. A. Lardon, B. V. Merkey, S. Martins, A. D\"otsch,
C. Picioreanu, J. U. Kreft, B. F. Smets, iDynoMiCS: next-generation individual-based 
modelling of biofilms, Environ. Microbiol. 13, 2416-34, 2011.
\bibitem{ibm_deformation} R. Sudarsan, S. Ghosh, J.M. Stockie, H.J. Eberl, Simulating biofilm deformation and detachment with the immersed boundary method, Communications in Computational Physics 19,  682--732, 2016.
\bibitem{Seminara}
A. Seminara, T.E. Angelini, J.N. Wilking, H. Vlamakis, S. Ebrahim, R. Kolter, D.A.
Weitz, M.P. Brenner, Osmotic spreading of Bacillus subtilis biofilms driven by an
extracellular matrix, Proc. Natl. Acad. Sci. USA 109, 1116--1121, 2012.
\bibitem{Picioreanu_surface} T. Storck, C. Picioreanu, B. Virdis, D.J. Batstone, Variable cell morphology approach for individual-based modeling of microbial communities, Biophysical 
Journal 106, 2037--2048, 2014.
\bibitem{Allen} M. A. A. Grant, B.Waclaw, R. J. Allen, P. Cicuta, The role of mechanical  forces in the planar-to-bulk transition in growing Escherichia coli microcolonies, J. R. Soc. Interface 11, 20140400, 2014.
\bibitem{poroelastic} A. Carpio, E. Cebri\'an, P. Vidal, Biofilms as poroelastic materials,  International Journal of Non-Linear Mechanics 109, 1--8, 2019.
\bibitem{solid/fluid} A. Carpio, E. Cebri\'an, Incorporating cellular stochasticity in solid-fluid mixture biofilm models, Entropy 22, 188, 2020.
\bibitem{ibm_adhesion} R. Dillon, L. Fauci, A. Fogelson,  D. Gaver, Modeling biofilm 
processes using the immersed boundary method, Journal of Computational Physics 129,  57--73, 1996.
\bibitem{ibm_rheology} J.A. Stotsky, J.F. Hammond, L. Pavlovsky, E.J. Stewart, J.G. Younger, M.J. Solomon, D.M. Bortz, Variable viscosity and density biofilm simulations using an immersed boundary method, Part II: Experimental validation and the heterogeneous rheology-IBM,
Journal of Computational Physics 317, 204--222, 2016.
\bibitem{ibm_division} Y. Li, A. Yun, J. Kim, An immersed boundary method for simulating a  single axisymmetric cell growth and division, Journal of Mathematical Biology 65, 653--675, 2012.
\bibitem{ibm_tumor} K.A. Rejniak,
An immersed boundary framework for modelling the growth of individual cells: an application to  the early tumour development, Journal of Theoretical Biology 247,  186--204, 2007.
\bibitem{ibm_multicellular} R. Dillon, M. Owen, K. Painter,
A single-cell-based model of multicellular growth using the immersed boundary 
method, In Moving Interface Problems and Applications in Fluid Dynamics (pp. 1--16). (Contemporary Mathematics). American Mathematical Society, 2008.
\bibitem{Hera} L. Chai, H. Vlamakis, R. Kolter, Extracellular signal regulation of cell differentiation  in biofilms, MRS Bulletin 36, 374--379, 2011.
\bibitem{Peskin95} C.S. Peskin, D.M. McQueen, A general method for the computer simulation  of biological systems interacting with fluids, Symposia of the Society for Experimental Biology 49, 265--76, 1995.
\bibitem{Peskin02} C.S. Peskin, The immersed boundary method, Acta Numerica 11, 479--517, 2002.
\bibitem{Deb18} B. Birnir, A. Carpio, E. Cebri\'an, P. Vidal, Dynamic energy budget approach to evaluate antibiotic effects on biofilms, Commun Nonlinear Sci Numer Simulat 54, 70--83, 2018.
\bibitem{DebBook} S.A.L.M. Kooijman, Dynamic energy budget theory for metabolic organization, Cambridge University Press 2008.
\bibitem{Debparameter}  T. Klanjscek, R.M. Nisbet, J.H. Priester, P.A. Holden,
Modeling physiological processes that relate toxicant exposure and bacterial population dynamics, PLoS One 7(2), e26955, 2012.
\bibitem{Youngcell1} H.H. Tuson, G.K. Auer,  L.D. Renner, M. Hasebe, C. Tropini, M. Salick, W.C. Crone, A. Gopinathan, K.C. Huang, D.B. Weibel, Measuring the stiffness of bacterial cells from growth rates in hydrogels of tunable elasticity, Mol Microbiol 84, 874--891, 2012.
\bibitem{Viscosity} K. Symon, Mechanics (Third edition), Addison-Wesley, 1971, ISBN 0-201-07392-7.
\end{thebibliography}
\end{document}